\documentclass[12pt]{article}

\usepackage{amssymb}

\newcommand {\union} {\cup}
\newcommand {\ZZ} {\mathbb Z}

\newcommand {\fI} {{\mathfrak{I}}}
\newcommand {\fX} {{\mathfrak{X}}}
\newcommand {\bQ} {{\mathbf{Q}}}
\newcommand {\bD} {{\mathbf{D}}}
\newcommand {\bX} {{\mathbf{X}}}
\newcommand {\CC} {{\mathbf{C}}}
\newcommand {\cJ} {{\mathcal{J}}}
\newcommand {\cV} {{\mathcal{V}}}

\newcommand {\secr} [1] {\S \ref{sec:#1}}

\newcommand {\RR} {{\mathbb  R}}
\newcommand {\subsubsubsection} [1] {\paragraph {#1}}

\usepackage {latexsym}
\newcommand {\pa} {\partial}
\newcommand {\id} {\mathbb I}

\newcommand {\tr} {\mbox {Tr}}

\newcommand {\beq} {\begin {equation}}
\newcommand {\eeq} {\end {equation}}

\newcommand {\beqn} {\begin {displaymath}}
\newcommand {\eeqn} {\end {displaymath}}

\newcommand {\beqar} {\begin {eqnarray}}
\newcommand {\eeqar} {\end {eqnarray}}

\newcommand {\beqarn} {\begin {eqnarray*}}
\newcommand {\eeqarn} {\end {eqnarray*}}

\newcommand {\nono} {\nonumber \\ {}}

\newcommand {\bary} {\begin {array}}
\newcommand {\eary} {\end {array}}

\renewcommand {\cal}[1] {\mathcal {#1}}

\newcommand {\half} {\frac 1 2}

\newcommand {\csp} {\hspace{5mm}}

\newcommand {\eqr} [1]  {{equation (\ref {eq:#1})}}

\newcommand {\Eqr} [1]  {{Equation (\ref {eq:#1})}}

\newcommand {\ignoretext} [1] {}

\renewcommand {\eqr} [1]  {{ (eq. \ref {eq:#1})}}
\renewcommand {\Eqr} [1]  {{(Eq. \ref {eq:#1})}}

\newcommand {\BSk}[1] [{ }] {\left|\left.#1\right>\!\right)}

\newcommand {\nod}[1] {: {#1} :}

\newcommand {\myI} [1] {\int \! #1 \,}

\newcommand{\be}{\begin{equation}}
\newcommand{\bea}{\begin{eqnarray}}
\newcommand{\eea}{\end{eqnarray}}
\newcommand{\ba}{\begin{array}}
\newcommand{\ea}{\end{array}}
\newcommand{\ee}{\end{equation}}

\newcommand{\vp}{{\varphi}}

\newcommand{\bj}{{\bar{j}}}

\makeatletter
\@addtoreset{equation}{section}

\usepackage {hyperref}

\begin{document}

\begin{titlepage}
\hfill
\vbox{ \footnotesize 
\halign{#\hfil     \cr   
           CALT-68-2410 \cr
           CERN-TH/2002-317 \cr
           TAUP-2717-02 \cr
           hep-th/0211147  \cr
           } 
      }  
\vspace*{4mm}
\begin{center}
{\Large {\bf  Families of $N=2$ Strings}\\}
\vspace*{10mm}
{\sc Yeuk-Kwan E. Cheung,}$^{a}$
\footnote{e-mail: {\tt cheung@theory.caltech.edu}}
{\sc Yaron Oz}$^{b\,c}$
\footnote{e-mail: {\tt yaronoz@post.tau.ac.il, Yaron.Oz@cern.ch}}
and {\sc Zheng Yin}$^{b}$
\footnote{e-mail: {\tt Zheng.Yin@cern.ch}}

\vspace*{4mm}
{\it{$^{a}$Theory Group, M.C. 452-48 \\
California Institute of Technology \\
Pasadena, CA 91125, U. S. A.}}  \\

\vspace*{4mm}
{\it {$^{b}$Theory Division, CERN \\
CH-1211 Geneva  23, Switzerland}}\\

\vspace*{4mm}
{\it {$^{c}$ Raymond and Beverly Sackler Faculty of Exact Sciences\\
School of Physics and Astronomy\\
Tel-Aviv University , Ramat-Aviv 69978, Israel}}\\
\vspace{5mm}

\end{center}

\begin{abstract}
\noindent 
In a given 4d spacetime bakcground, one can often construct not one but
a family of distinct $N=2$ string theories.   This is due to the
multiple ways an $N=2$ superconformal algebra can be embedded in a given
worldsheet theory.  We formulate the principle of obtaining different
physical theories by gauging different embeddings of the same symmetry
algebra in the same ``pre-theory.''   We then apply it to the $N=2$ strings
and formulate the recipe for finding the associated parameter
spaces of gauging.  Flat and curved target spaces of both $(4,0)$ and
$(2,2)$ signatures are considered.  We broadly divide the gauging
choices into two classes, denoted by $\alpha$ and $\beta$,
and show them to be related by T-duality.  The distinction between them
is formulated topologically and hinges on some unique properties of 4d
manifolds.  We determine what their parameter spaces of gauging are
under certain  simplicity ansatz for generic flat spaces ($\RR^4$ and
its toroidal compactifications) as well as some curved spaces. We
briefly discuss the spectra of D-branes for both $\alpha$ and $\beta$ 
families.
\end{abstract}
\end{titlepage}

\tableofcontents

\newpage

\section{Introduction and Summary of Results}     
 
The theories that have come to be known as $N=2$ strings \cite{Ademollo:1976pp}
date back to the period of transition of string theory from 
a candidate theory of hadrons to that of grand unification \cite{Scherk:1974ca}.
The worldsheet theory of $N=2$ string has
gauged $N=2$ superconformal symmetry, which explains 
its name\footnote
    {To be precise, the superconformal symmetry is really $(2,2)$.  
    However, since throughout this paper we shall only consider 
    theories with the same amount of supersymmetries for the 
    left and right movers, we shall simply write $N=2$ without 
    ambiguity.  Similar conventions apply to $N=1$ and $N=4$.}.
Compared with $N=1$ superconformal field theory, 
which gives rise to superstring theory, it adds no 
new kind of fields, but another supersymmetry and an $U(1)$ 
R-symmetry, for the left and right movers independently.
The resulting $N=2$ superconformal symmetry is 
all gauged.
This theory can be thought as gauge fixed version of $N=2$ gauged 
supergravity in two dimensions.

Progress in this theory has seen fits and starts ever since.  
Deep connection has been found between 
the closed $N=2$ strings 
\footnote{In this work we exclusively consider critical $N=2$ string, so 
    we will drop the qualifier ``critical.''
    The matter part of the worldsheet conformal field theory has 
    central charge $6$.  The target space dimension of the corresponding 
    $\sigma$-model is $4$.}
and 
the self-dual solution of the Einstein equations \cite{Ooguri:1991fp}, 
between the open $N=2$ strings and
self-dual Yang-Mills \cite{Marcus:1992xt}, and between 
heterotic $N=2$ strings and reduction/deformation of 
self-dual Yang-Mills \cite{Ooguri:1991ie}.                    
A relation between the
$N=2$ strings and integrable models was also suggested in
\cite{Ooguri:1991fp}.  These have been the cause for its continuing
fascination and alone justify a thorough study.

Unfortunately in
reality studies of $N=2$ strings have been somewhat disconnected 
from $N=1$ superstrings, not the least because the former propagates 
in space with an even number of temporal dimensions.  Yet there is a 
compelling feature of $N=2$ strings that should appeal 
even to those who exclusively study superstrings:
its spacetime 
is relative conventional, including flat
$\RR^4$ as the simplest case \footnote{The $N=4$ string has been
argued to be equivalent to the $N=2$ string in \cite{Siegel:1992ev}.}.
Unlike strings with 
less worldsheet spacetime supersymmetry, i.e. superstrings and
bosonic strings, $N=2$ strings typically have only finitely many degrees of
freedom.  In fact in the simplest setting, $\RR^4$ of signature $(2,2)$, 
it has long been established that
only one scalar field propagates \cite{Ademollo:1976pp}.  
This makes it the simplest string
theory with some physical, if not directly phenomenological, meaning.  
As is well-known, the infinite number of fields propagating in
superstring (or bosonic string) theories render it extremely difficult to
carry out  analysis standard in quantum field theory 
unless one first takes a drastic limit that only keeps a finite numbers of 
fields.  This often confines brilliant ideas to a
state of conjectures rudimentarily tested in special limits.
The simplicity of
$N=2$ strings suggests itself as a promising theoretical guinea pig
where careful analysis and definite results can be obtained in a
robust manner.  That could be of great help to the study of superstrings
or indeed any theories of strings.
$N=2$ strings seem to have no spacetime supersymmetry
\footnote{See, however, \cite{Chalmers:2000bh}.}.
Issues such as the moduli fixing problem and
the resolution of space-time singularities may be addressed here in a 
simplified
set-up.

A concrete step in this program would be to understand better 
what all the
possible $N=2$ strings are.  More than one kind of $N=2$ string 
theories have been known to exist, 
but when we looked into this 
it came as a surprise that the possibilities of different $N=2$ strings 
had not been systematically
studied, characterized, or understood.  
Important insights into the geometry of the $N=2$ string was
obtained by Ooguri and Vafa in \cite{Ooguri:1991fp} where they showed
that the physical scalar field of the $N=2$ string theory describes
deformation to the  K\"{a}hler potential of a target space that is K\"{a}hler and the
equation of motion at the tree level requires the metric be
hyper-K\"{a}hler.\footnote{ 
  It is in this sense that the $N=2$ string
  has been called a theory of self-dual gravity.  This name is
  historical but somewhat unfortunate because among other things, 
  only deformations of the K\"{a}hler potential with respect to a choosen complex structure is a 
  fluctuating quantum fields in the 
  theory, even though many other deformations of the metric 
  can also maintain the self-duality of 
  the Riemann tensor. }
However, this cannot be the only possible $N=2$ theory, even in $\RR^4$.
To see this, let us perform for the $N=2$ string a standard procedure:
compactify the $N=2$ string on a circle of radius $R$.  Like any
other string theory there is a T-dual interpretation in which the
spacetime involves a circle of radius $\frac{\alpha'}{R}$.  If we take
$R$ to infinity, we recover in the original description the $N=2$
string propagating in $\RR^4$ as discussed in \cite{Ooguri:1991fp}.  We
shall call this theory $\beta$-string.  If instead we take $R$ to
zero, the T-dual space becomes $\RR^4$ and we again have an $N=2$ string
theory propagating in $\RR^4$, which we shall call $\alpha$-string.
However, $\alpha$ and $\beta$ are not the
same: they have different 3-point amplitudes \cite{Gluck:2003wg}.

In this paper we undertake a systematic classification of $N=2$
strings propagating in large class of target spaces.  Our point of
departure is fundamental: 
any $N=2$ string worldsheet theory, in the covariant
formalism of the conformal field theory(CFT) approach, is obtained by
gauging an $N=2$ superconformal algebra(SCA) respectively
for the left and right moving chiral parts of the CFT.
Therefore the physical theory is controlled both by the choice of $(2,2)$ 
superconformal field, with central charge $c=6$ to cancel the 
superconformal anomaly of the BRST ghosts, and by the choice of a 
$(2,2)$ SCA within that CFT to gauge.  The second choice is far from 
unique, because many known examples of $c=6$ $N=2$ superconformal field 
theories (SCFT) have in fact 
$N=4$ SCA which allows a continuum of embeddings of $N=2$ SCA.  
However, different embeddings are actually physically equivalent 
if and only if they are related by some symmetry of the CFT.  
Therefore one must take care to quotient the choices of $N=2$ embeddings 
by the action of the field theory symmetries.
To make a connection with spacetime, 
one considers a $\sigma$-model\footnote{
         In this work we only consider \emph {nonlinear}
         $\sigma$-models so we will refer to them simply as 
         $\sigma$-models.}
realization of the $N=2$ SCFT.  
Different SCFTs correspond to different spacetimes whose geometry 
satisfies  the $\beta$-functional equations as well as conditions 
for worldsheet supersymmetries.  
The choice of the left and right chiral $N=2$ SCAs 
corresponds to choosing two complex structures.  Under certain simplifying
assumptions, we find that the $N=2$ string theories can be broadly
divided into two types, which can be interpreted in a well defined way
as the generalization of the $\alpha$ and $\beta$ theories in $\RR^4$.  
This follows a careful analysis of the space of complex structures in 
$\RR^4$ and the distinction is topological.  It applies to curved spaces 
as well as flat spaces.
In general, $\alpha$ and $\beta$ strings cannot propagate in the same 
spaces, but flat spaces are notable and fortunate exceptions.
Depending on the details of the target space geometry, there are
in addition a number of continuous parameters with each type.  
Some of the parameters are well-known --- they 
control the $N=2$ moduli of the SCFT and correspond to such geometric 
quantities as the sizes and shapes of the target space.  
We are concerned in this work with the rest, which 
specify the selection of an embedding of the left and right $N=2$ SCAs 
in a given CFT.  They are the \emph {parameters of gauging}.
We study carefully all generic flat spaces
of both Euclidean and $(2,2)$ signatures
and consider some prominent examples of curved spaces.

The simplifying assumptions just mentioned were made to enable a
rigorous and exhaustive analysis.  They are clearly and precisely
stated in \eqr{SimpleNTwoRealization} and
\eqr{SimpleEmbeddingNTwoInNFourCompact}.  Relying on them means that we
cannot prove or claim that there are no other possibilities of the $N=2$
strings.  Nonetheless, we wish to stress that these assumptions 
are based on very reasonable geometric expressions.  
Other possibilities, if any, would be more
intricate.  If they do exist, they are not likely to blur the 
above mentioned board division into 2 types, which is based 
on topological considerations.
In the past literature, there had been scattered 
examples $N=2$ string theories beyond the one studied in
\cite{Ooguri:1991fp}.  They turned out to be particular cases in
the classification scheme we derive in this paper.  We comment on a
few of them here.  \cite{Berkovits:1994vy} suggested that the
$SU(2)$ outer automorphism of $N=4$ SCA can be used to obtain different
$N=2$ strings.  This would roughly correspond to 
$\beta$-strings propagating on a generic $K3$, which is analyzed 
in \secr{KThree} of  in the present paper.  
\cite{Gates:1988tn} and \cite{Grisaru:1997pg} considered
employing elaborate variations of superspace formalism to study $N=2$
strings.  Their theories are examples of $\alpha$ and $\beta$ types.  Our unifying approach brings these examples and beyond 
together and organize them based on a scheme derived from the basic 
principles of $N=2$ strings. Formulated solely in terms of target space 
geometry, it does not assume or rely on 
any particular $N=2$ superspace representations.  Of course, the latter 
can be advantageous for calculations in specific cases.

The original impetus for us to approach  $N=2$ strings was to study
D-branes in the theory, in light of late developments in the $N=1$
superstrings.  It soon proved necessary to understand the closed
string theory better and thus we steered toward the current study.  
Nonetheless D-branes are relevant to the classification of
closed string theories.  We have found that just as in the type IIA
and type IIB superstrings, the $N=2$ $\alpha$ and $\beta$ strings
admit D-branes of different dimensionalities.  In this paper we
mention briefly these D-brane results, leaving a full analysis to 
\cite{Gluck:2003pa}.
 
\subsection{Guide to the paper}

In section 2 we discuss in details the relevant aspects of flat four
dimensional spaces with signatures $(4,0)$ and $(2,2)$. These will be
used in the following sections of the paper. For those who are
too impatient to read the whole section, 
the most important results are summarized at its beginning. 
In section 3 we review the facts of two-dimensional $N=2$ and $N=4$
compact and noncompact superconformal algebras (SCA) and their
spacetime meaning.  We also illustrate some embeddings of $N=2$ SCA in
$N=4$ SCA. They are needed later for the study of $N=2$
strings in curved spaces.  People who are already familiar with the compact $N=4$ 
SCAs can skip
this section and return later for a quick glance at the noncompact
$N=4$ SCA in \secr{NoncompactNFour} as well as \secr{FatFreeNIsFour},
where we present a simple extension of the $N=4$ SCAs that is tailored
to the free field realization by the $\sigma$-model on $\RR^4$.  
In section 4 we will study the families of $N=2$
strings on $\RR^4$ and $\RR^{2,2}$  as well as their toroidal
compactifications.  We propose an ansatz of
simple realization of $N=2$ SCA and find all solutions.  They fall into
two broad types, corresponding to the $\alpha$ and $\beta$ strings
mentioned above.  We also study the continuous parameters.  Then we
generalize the notion of the $\alpha$-$\beta$ divide to general spacetimes
with a precise  topological definition. 
Section 5 concentrates on the study of curved spacetime backgrounds.  A
general methodology is given based on an ansatz of simple embedding. 
We discuss the specific 
case of $K3$ and outline the steps needed to study other
cases, such as hyper-K\"{a}hler spaces with isometries.
In section 6 we study the behavior of the $N=2$ strings under
T-duality and shows in particular that $\alpha$ and $\beta$ strings can
be related in this way.
In section 7 we 
briefly discuss the D-branes of the $N=2$ strings. 
In the appendix we 
summarize the notations and conventions adopted in this paper.
 
\subsection {Summary of Results}

The calculational results  are summarized in the  following tables.  

\begin{table}
\centering
\begin{tabular}  {||c|l|c|c||}  \hline
Family & $\fX$ & Rotational Isometry & Parameter Space of Gauging\\ \hline
 & $\RR^4$ & $SU(2)\times SU(2)'$ & 1 point\\ \cline{2-4}
 & $\RR^3 \times S^1$ & $O(3) \times \fI$ & $(RP^2\times RP^2)/SU(2)$\\ \cline{2-4}
$\alpha$ & $\RR^2\times T^2$ & $O(2) \times \fI$ & $(RP^2\times RP^2)/U(1)$ \\ \cline{2-4}
 & $\RR\times T^3$ & $\ZZ_2 \times \fI$ & $RP^2\times RP^2$\\ \cline{2-4}
 & $T^4$ & $\fI$ & $RP^2\times RP^2 \union RP^2\times RP^2$\\ \hline
 & $\RR^4$ & $SU(2)\times SU(2)'$ & $(RP^2\times RP^2)/SU(2)$\\ \cline{2-4}
 & $\RR^3\times S^1$ & $O(3) \times \fI$ & $(RP^2\times RP^2)/SU(2)$ \\ \cline{2-4}
$\beta$ & $\RR^2\times T^2$ & $O(2) \times \fI$ & $(RP^2\times RP^2)/U(1)$ \\ \cline{2-4}
 & $\RR\times T^3$ & $\ZZ_2 \times \fI$ & $RP^2\times RP^2$ \\ \cline{2-4}
 & $T^4$ & $\fI$ & $RP^2\times RP^2 \union RP^2\times RP^2$ \\ \cline{2-4}
 & K3 & $\emptyset$ & $RP^2\times RP^2 \union RP^2\times RP^2$ \\ \hline
\end{tabular}
\caption{Families of N=2 Strings in flat Euclidean and generic K3 backgrounds.}
\end{table}

\begin{table}
\centering
\begin{tabular}  {||c|l|c|c||}  \hline
Family  & $\fX$ & Rotational Isometry & Parameter Space of Gauging\\ \hline              & $\RR^{(2,2)}$        & $SL(2)\times SL(2)'$  & 1 point  \\  \cline{2-4}              & $\RR^{(2,1)}\times S^1$ & $O(2,1) \times \fI$    & $(S^2_{1+}\times S^2_{1+})/SL(2)$ \\ \cline{2-4} $\alpha$ 
        & $\RR^{(2,0)}\times T^{(0,2)}$         &$O(2) \times \fI$ &  $(S^2_{1+}\times S^2_{1+})/U(1)$   \\ \cline{2-4} 
                & $\RR^{(1,1)}\times T^{(1,1)}$         &$O(1,1) \times \fI$       &  $(S^2_{1+}\times S^2_{1+})/SO(1,1)$  \\ \cline{2-4}           & $\RR\times T^{(1,2)}$                 &$\ZZ_2 \times \fI$            &  $S^2_{1+}\times S^2_{1+}$  \\ \cline{2-4} 
                &$T^{(2,2)}$                            &$\fI$            &  $S^2_{1+}\times S^2_{1+} \union S^2_{1+}\times S^2_{1+}$  \\   \hline               & $\RR^{(2,2)}$         &$SL(2)\times SL(2)' $&
  $(S^2_{1+}\times S^2_{1+})/SL(2)$ 
   \\ \cline{2-4}               & $\RR^{(2,1)} \times S^1$ &$O(2,1) \times \fI$     
&  $(S^2_{1+}\times S^2_{1+})/SL(2)$   \\ \cline{2-4} $\beta$   
&$\RR^{(2,0)}\times T^{(0,2)}$  &$O(2) \times \fI$  &  $(S^2_{1+}\times S^2_{1+})/U(1)$  \\ \cline{2-4} 
                &$\RR^{(1,1)}\times T^{(1,1)}$  &$ O(1,1) \times \fI$ &  $(S^2_{1+}\times S^2_{1+})/SO(1,1)$          \\ \cline{2-4} 
                & $\RR\times T^{(1,2)}$                 &$ \ZZ_2 \times \fI$            & $S^2_{1+}\times S^2_{1+}$  \\ \cline{2-4} 
                &$T^{(2,2)}$                    &$\fI$            & $S^2_{1+}\times S^2_{1+} \union S^2_{1+}\times S^2_{1+}$ \\  \hline
\end{tabular}
\caption{Families of N=2 Strings in Flat (2,2) Backgrounds.}
\end{table}

\section {Complex Geometries and Symmetries of $\RR^4$}     
\label {sec:GeometryOfRFour}

In this work we consider exclusively critical $N=2$ strings, therefore
the target spacetime is four dimensional.  In this section we derive
some interesting properties  of $\RR^4$ which we will use throughout
the paper.  They apply globally to any 
flat target spaces and locally to the tangent bundle of each point on a curved target spaces.  We consider the case
of the metric signature $(4,0)$ and $(2,2)$ separately in great details.

The most important results on this section is summarized here for the
impatient readers: 
\begin {quote}
Given an positive definite metric in $\RR^4$, the space of complex
structures with respect to which the metric is Hermitian consists of
two disjoint
2-spheres. 
Given a metric of $(2,2)$ signature, that space consists of two
disjoint hyperboloids of two sheets.  
In both cases, the two components are interchanged by a parity changing 
rotation.
\end{quote}
The proof can be found in 
\secr{HermitianStructureEuclidean} and \secr{HermitianStructureTwoTwo}

\subsection {Euclidean Spaces}

\subsubsection {Vector and Bi-spinor Notation}  
\label {sec:SPINFourToSUTwoSUTwo}
The symmetry group leaving the metric invariant is $O(4)$.  Let $V$ denote
a 4 component real vector with entries $V^I$.  The vector representation of
$O(4)$ is realized as the group of linear transformation leaving invariant
the quadratic form
\beq
        \left<V, V \right> \equiv \cal G_{IJ} V^I V^J
\eeq
the components of an element of its vector representation.  Without loss of
generality we shall set the metric to be equal to the identity matrix, so
$\left<V, V \right>$ is just the sum of the squares of $V^I$.  Define now
\beqar 
\label {eq:BispinorFromVectorEuclidean}
        {\cal V^\alpha}_\beta &\equiv& 
                        {(V^0 \id + \imath V^i {\sigma^i})^\alpha}_\beta  \nono
                &=& \left( \bary {rr} V^0 + \imath V^3 & V^2 + \imath V^1\\ 
        -V^2 + \imath V^1 & V^0 - \imath V^3 \\ \eary \right) \ .
\eeqar
Therefore
\beq 
\det {\cal V} = \left< V, V \right> \ .
\eeq 
The inverse relation is 
\beq    
\label {eq:VectorFromBispinorEuclidean}
        V^0 = \half \tr (\cal V), \csp 
        V^i = \frac {- \imath} 2 \tr (\cV \sigma^i) \ .
\eeq
Because $V^I$ is are real, 
\beq    \label {eq:EuclideanRealityConditionForCalV}
         ({\cal V^\alpha}_\beta)^* = {(\sigma^2)^\alpha}_{\alpha'} 
         {\cal V^{\alpha'}}_{\beta'} {(\sigma^2)^{\beta'}}_{\beta}.
\eeq
Now consider the linear transformation 
\beq \label {eq:GroupActionOnBispinor}
        \cal V \to U \cal V U'^{-1} \ .  
\eeq 
We want to maintain the reality condition 
\eqr{EuclideanRealityConditionForCalV} and keep $\det \cal V$ invariant.  
This can be arranged if
\beqar 
        \sigma^2 U \sigma^2 &=& U^*, \nono \det U &=& 1,
\eeqar 
and the same for $U'$.  This means none other than that $U$ and $U'$
are $SU(2)$ matrices
\beq
        \exp(\imath \Lambda_i \sigma^i).
\eeq
Yet there is no condition relating $U$ and $U'$.  To distinguish the action of the two $SU(2)$'s we shall call the one acting on the second indices $SU(2)'$.  We have here a realization of the $(2,2)$ representation of $SU(2)\times SU(2)'$ within the vector representation of $O(4)$.  This is hardly surprising as
\beq
        SU(2) \times SU(2) = SPIN(4) \ .
\eeq
\Eqr {BispinorFromVectorEuclidean} and \eqr {VectorFromBispinorEuclidean} then let us convert between the vector and the bi-spinor form of the same representation.

\subsubsection {Parity}        
\label {sec:ParityOperationEuclidean}
The overlap of $SPIN(4)$ and $O(4)$ is $SO(4)$.  The center in $SPIN(4)$
that is trivial in $SO(4)$ is $U = U' = - \id$.  The component in $O(4)$
that is missing from $SO(4)$ is generated from the latter by any parity
operation: reflection of an odd number of directions, e.g. $V^i \to - V^i$. 
The latter has a simple manifestation in $\cal V$.  Consider
\beq
        {{\cal V'}^\alpha}_\beta = \epsilon^{\alpha\alpha'} 
        {\cal V^{\beta'}}_{\alpha'} \epsilon_{\beta'\beta}   \ .
\eeq
Since the $\epsilon$ tensor used here is $SU(2)$ invariant, ${\cal V'}$
transform the same way as $\cal V$ and satisfies the same reality condition
\eqr{EuclideanRealityConditionForCalV}: it is an alternative form of
writing a vector in terms of bi-spinor.  In fact,
\beq
        {\cal V'} 
        = V^0 - \imath V^i \sigma^i
\eeq
so the inversion of $V^i$ interchanges $\cal V$ and ${\cal V'}$.  Alternatively, define
\beq    
\label {eq:CalVDownDown}
        {\cal V}^{\alpha,\beta} 
        = {\cal V^{\alpha}}_{\beta'}\epsilon^{\beta'\beta} \ .
\eeq
Then the same operation is realized by swapping the two spinor indices:
\beq
        {\cal V}^{\alpha,\beta} \leftrightarrow {\cal V}^{\beta,\alpha} \ .
\eeq
Compose this with $SU(2)\times SU(2)$ has the effect of exchanging the two
$SU(2)$'s.  We shall often use the representation\eqr{CalVDownDown} in
this work.  It transforms under $ SU(2)\times SU(2)$ as
\beq
        \cal V^{\alpha,\beta} \to 
        {U^\alpha}_{\alpha'} {U'^\beta}_{\beta'} \cal V^{\alpha',\beta'}
\eeq
and satisfies the  reality condition 
\beq    \label {eq:RealityConditionForCalVUpUpEuclidean}
        (\cal V^{\alpha,\beta})^* = 
        - {(\sigma^2)^\alpha}_{\alpha'} {(\sigma^2)^\beta}_{\beta'} 
                \cal V^{\alpha'\beta'} \ .
\eeq
Since 
\beq
        \det {{\cal V^\alpha}_\beta} 
        = \det {\cal V^{\alpha,\beta}} = \half 
        \epsilon_{\alpha\alpha'} \epsilon_{\beta\beta'} 
        \cal V^{\alpha,\beta} \cal V^{\alpha',\beta'}
\eeq
the metric in the form of tensor product of bi-spinors is 
\beq    
\label {eq:MetricInBispinorForm}
         \half \epsilon_{\alpha\beta} \otimes \epsilon_{\alpha'\beta'} \ .
\eeq

\subsubsection {Complex Structures}    
\label {sec:HermitianStructureEuclidean}
An almost complex structure ${\cal J^I}_J$ is a real matrix satisfying 
\beq
        \cal J \cal J = - \id \ .
\eeq
It is said to make a metric $\cal G$ Hermitian if 
\beq
        \cal J^\top \cal G \cal J = \cal G
\eeq
which implies that $\cal K \equiv \cal G \cal J$ is antisymmetric
\beq
        \cal J^\top \cal G = - \cal G \cal J \ .
\eeq

Antisymmetric $4\times 4$ matrices span a 6 dimensional space.  A set of
basis for them in bi-spinor form and satisfying the reality condition \eqr{RealityConditionForCalVUpUpEuclidean} can be conveniently written in terms
of tensor product:
\beqar    \label {eq:PairOfSUTwoKahlerForm}
        2\cal K^{[i]}_{\alpha,\alpha';\beta,\beta'} 
        &=& - \imath (\sigma^i_{\alpha\beta}) 
        \otimes (\epsilon_{\alpha'\beta'}), \nono 
        2\cal K'^{[i]}_{\alpha,\alpha';\beta,\beta'} 
        &=& - \imath (\epsilon_{\alpha\beta}) 
        \otimes (\sigma^i_{\alpha'\beta'}) \ .
\eeqar
  From $\cal K=\cal G \cal J$ one finds the corresponding $\cal J$.  They are 
\beqar    \label {eq:PairOfSUTwoComplexStructure}
        {\cal (J^{[i]})^{\alpha,\alpha'}}_{\beta,\beta'} 
        &=& - \imath ({(\sigma^i)^\alpha}_\beta) 
        \otimes ({\delta^{\alpha'}}_{\beta'}) \ , \nono 
        {(\cal J'^{[i]})^{\alpha,\alpha'}}_{\beta,\beta'} 
        &=& - \imath ({\delta^\alpha}_{\beta}) 
        \otimes ({(\sigma^i)^{\alpha'}}_{\beta'}) \ .
\eeqar
It is clear that they transform in the adjoint representations of
$so(3)$ and $so(3)'$ respectively.  A parity operation exchanges them
as it exchanges the two $SU(2)$'s.  Furthermore, they satisfy the
following algebraic relations:
\beq    
\label {eq:HomogeneousRelationOfComplexStructureJSUTwo}
         [ {\cal J}^{[i]}\ , \ \cal J'^{[j]} ] = 0 \ , 
\eeq
and 
\beqar  
\label {eq:AlgebraicRelationOfComplexStructureJSUTwo}
        {\cal J}^{[i]}  {\cal J}^{[j]} 
        &=& - \delta^{ij} + \epsilon^{ijk} {\cal J}^{[k]}\ , \nono
        \cal J'^{[i]}  \cal J'^{[j]} &=& - \delta^{ij} +\epsilon^{ijk} \cal J'^{[k]} \ .
\eeqar
Therefore they themselves are generators of $SU(2)\times SU(2)'$ in the
$(\bf{2},\bf{2})$ representation. The nine anti-commutators $\{\cal J^{[i]}, \cal
J'^{[j]}\}$ do not vanish.  

An arbitrary $4\times 4$ antisymmetric matrix $\cal K = \cal G \cal J =
\cal G \sum_i (a_i \cal J^{[i]} + a'_i \cal J'^{[i]})$.  The condition $\cal J
\cal J = - \id$ dictates that
\beq
        \sum_i (a_i)^2 + (a'_i)^2 = 1, \csp
        a_i a'_j = 0
\eeq
so the only solutions are either 
\beq    \label {eq:JOnSphere}
        \cal J = \sum_i a_i\cal J^{[i]}, \csp \sum_i (a_i)^2 = 1 
\eeq
or
\beq
        \cal J = \sum_i a'_i {\cal J'}^{[i]}, \csp \sum_i (a'_i)^2 = 1 \ .
\eeq
Thus they are located on two disjoint spheres which are exchanged by parity operations.

\subsubsection {Stabilizer}
\label {sec:StabilizerEuclidean}
In this course we shall often have to consider subgroup of the orthogonal
group leaving invariant some/several subspaces of $\RR^4$.

\subsubsubsection{Stabilizer of $V^0$}
Clearly, an $O(3)$ subgroup of the $O(4)$ symmetry leaves $V^0$ invariant. 
It is equally clear that a diagonal subgroup of $SU(2)\times SU(2)$,
\beq
        U U' = \id
\eeq
leaves $V^0$ invariant in \eqr {GroupActionOnBispinor}.  This accounts for the $SO(3)$.  The full $O(3)$ is obtained by considering an parity operation such as $V^i \to - V^i$.  It corresponds to $\cal{V} \to - {\cal{V'}}$, or $\cal V_{\alpha\beta} \to -\cal{V}_{\beta\alpha}$

\subsubsubsection {Stabilizer of $V^0$ and $V^3$}
The obvious $O(2)$ subgroup leaving $V^0$ and $V^3$ invariant is, in the bi-spinor form, generated by the $U(1)$
\beq
        e^{\imath \theta \sigma^3}
\eeq
plus a reflection on, say, $V^2$.

\subsection {$(2,2)$ Space}

\subsubsection {Vector and Bi-spinor Notation}
 \label {sec:SPINTwoTwoToSLTwoSLTwo}
The symmetry group leaving the metric invariant is $O(2,2)$.  Let $V$ denote a 4 component real vector with entries $V^I$.  The vector representation of $O(4)$ is realized as the group of linear transformation leaving invariant the quadratic form
\beq
        \left<V, V \right> \equiv \cal G_{IJ} V^I V^J
\eeq
the components of an element of its vector representation.  Without loss of generality we shall set the metric to be such that
\beq
        \left<V, V \right> = V^0 V^0 + V^2 V^2 - V^1 V^1 - V^3 V^3 \ .
\eeq
The relation with Euclidean space is that we have $V^1$ and $V^3$ analytically continued to the imaginary axis.  Define now
\beqar    \label {eq:BispinorFromVectorTwoTwo}
        {\cal V^\alpha}_\beta &\equiv&
  {(V^0 \id + V^1 {\sigma^1} 
  + \imath V^2 {\sigma^2} + V^3 {\sigma^3})^\alpha}_\beta \nono
        &=& \left( \bary {rr} V^0 + V^3 & V^2 + V^1\\ 
        -V^2 + V^1 & V^0 - V^3 \\ \eary \right) \ .
\eeqar
Because $V^I$ are real, 
\beq    \label {eq:TransLorentzianRealityConditionForCalV}
        {\cal V}^* = \cal V \ ;
\eeq
and
\beq
        \det {\cal V} = \left< V, V \right> .
\eeq
Now consider the linear transformation 
\beq    \label {eq:GroupActionOnBispinorTwoTwo}
        \cal V \to U \cal V U'^{-1} \ .
\eeq
We want to preserve the reality condition 
\eqr {TransLorentzianRealityConditionForCalV} and keep $\det \cal V$ invariant.  This can be arranged if
\beqar
         U^* &=& U, \nono
        \det U &=& 1 \ ,
\eeqar
and the same for $U'$.  This demands that $U$ and $U'$ be $SL(2,\RR)$ matrices
\beq
        \exp(\Lambda_1 \sigma^1 + \Lambda_3 \sigma^3 + \imath \Lambda_2 \sigma^2)
\eeq
Yet $U$ and $U'$ are unrelated.  To distinguish the action of the two $SL(2,\RR)$'s we shall call the one acting on the second indices $SL(2,\RR)'$.  Hence we have found a realization of the $(2,2)$ representation of $SL(2,\RR)\times SL(2,\RR)$ within the vector representation of $O(2,2)$.   Again, this is hardly surprising since
\beq
        SL(2,\RR)\times SL(2,\RR) = SPIN(2,2) \ . 
\eeq
\Eqr {BispinorFromVectorTwoTwo} 
lets us convert between the vector and the bi-spinor forms of the same representation.

\subsubsection {Parity}  \label {sec:ParityOperationTwoTwo}
Unlike $SO(4)$, $SO(2,2)$ is not connected.  
Besides the component of $SO(2,2)$ containing the
identity, which we denote as $SO_0(2,2)$, or the ``proper'' $SO(2,2)$, there is also the ``improper'' one generated from $SO_0(2,2)$ by the reflection of one ``time'' and one ``space'' direction.  They correspond to $U$ and $U'$ both having determinant $-1$ and still being real.  An example is
\beq
        U = U' = \sigma^1 \ .
\eeq
It reverses the sign of $V^2$ and $V^3$. The overlap of $SPIN(2,2)$ and $O(2,2)$ is $SO_0(2,2)$.  The center in $SPIN(2)$ that is trivial in $SO(2,2)$ is $U = U' = - \id$.  We have just shown how to go from $SO_0(2,2)$ to $SO(2,2)$.  The two other components in $O(2,2)$ that is missing from $SO(2,2)$ are generated from the latter by any parity operation: reflection of an odd number of directions, e.g.  $V^i \to - V^i$.  The latter has a simple manifestation in $\cal V$.  Consider
\beq
        {{\cal V'}^\alpha}_\beta = \epsilon^{\alpha\alpha'} 
        {\cal V^{\beta'}}_{\alpha'} \epsilon_{\beta'\beta}   \ .
\eeq
Since the $\epsilon$ tensor used here is $SL(2,\RR)$ invariant,  $\cal V'$ transform the same way as $\cal V$ and satisfies the same reality condition\eqr{EuclideanRealityConditionForCalV}: it is an alternative form of writing a vector in terms of bi-spinor.  In fact,
\beq
        \cal V' 
        = V^0 - \imath V^i \sigma^i \ ,
\eeq
so the inversion of $V^i$ interchanges $\cal V$ and $\cal V'$.
Alternatively, define 
\beq    \label {eq:CalVDownDownTwoTwo}
        {\cal V}^{\alpha,\beta} 
        = {\cal V^{\alpha}}_{\beta'}\epsilon^{\beta'\beta}.
\eeq
Then the same operation is realized as swapping the two spinor indices:
\beq
        {\cal V}^{\alpha,\beta} \to {\cal V}^{\beta,\alpha} \ .
\eeq
Compose this with $SL(2,\RR)\times SL(2,\RR)$ has the effect of exchanging the
two $SL(2,\RR)$'s.  We shall often use the representation 
\eqr{CalVDownDownTwoTwo}
in this work.  It transform under $SL(2,\RR)\times SL(2,\RR)$ as
\beq
        \cal V^{\alpha,\beta} \to 
        {U^\alpha}_{\alpha'} {U'^\beta}_{\beta'} \cal V^{\alpha',\beta'}
\eeq
and has reality condition 
\beq    \label {eq:RealityConditionForCalVUpUpTwoTwo}
        (\cal V^{\alpha,\beta})^* = \cal V^{\alpha,\beta} \ .
\eeq
Since 
\beq
        \det {{\cal V^\alpha}_\beta} 
        = \det {\cal V^{\alpha,\beta}} = \half 
        \epsilon_{\alpha\alpha'} \epsilon_{\beta\beta'} 
        \cal V^{\alpha,\beta}\cal V^{\alpha',\beta'}
\eeq
the metric in the form of tensor product of bi-spinors is 
\beq    
        \half \epsilon_{\alpha\beta} \otimes \epsilon_{\alpha'\beta'} \ .
\eeq

Note that even though what differentiates $(2,2)$ from $(4,0)$ is the metric, 
in the  bi-spinor notation the forms of the metrics 
are actually the same as in \eqr{MetricInBispinorForm}. This is because 
\beq    \label {eq:NormSquaredIsDeterminant}
        \left< V, V \right> = \det {\cal V}
\eeq
holds for both cases.  The difference in the metric is instead reflected by the simpler reality condition\eqr{RealityConditionForCalVUpUpTwoTwo}.   This makes sense as the two are related by analytic continuation of $V^1$ and $V^3$  to the imaginary axis.  Everything simply is real as in \eqr{TransLorentzianRealityConditionForCalV}.

\subsubsection {Complex Structures}     
\label {sec:HermitianStructureTwoTwo}
An almost complex structure ${\cal J^I}_J$ is a real matrix satisfying 
\beq
        \cal J \cal J = - \id \ .
\eeq
It is said to the make the metric $\cal G$ Hermitian if 
\beq
        \cal J^\top \cal G \cal J = \cal G
\eeq
which implies that $\cal K \equiv \cal G \cal J$ is antisymmetric
\beq
        \cal J^\top \cal G = - \cal G \cal J \ .
\eeq

This reality condition \eqr{RealityConditionForCalVUpUpTwoTwo} 
implies that both $\cal K$ and $\cal J$ are real.
Real anti-symmetric $4\times 4$ matrices are spanned by the following 
set of basis with real coefficients:
\beqar  \label {eq:PairOfSLTwoKahlerForm}
        2 \cal K^{[i]}_{\alpha,\alpha';\beta,\beta'} 
        = (\sigma^i_{\alpha\beta}) 
        \otimes (\epsilon_{\alpha'\beta'}), \csp i = 1, 3; &\csp&
        2\cal K^{[2]}_{\alpha,\alpha';\beta,\beta'} 
        = ( -\imath \sigma^2_{\alpha\beta}) 
        \otimes (\epsilon_{\alpha'\beta'}) \ , \nono
        2\cal K'^{[i]}_{\alpha,\alpha';\beta,\beta'} 
        = (\epsilon_{\alpha\beta})
        \otimes (\sigma^i_{\alpha'\beta'}), \csp i = 1, 3; &\csp&
        2\cal K'^{[2]} = (\epsilon_{\alpha\beta}) 
        \otimes (- \imath \sigma^2_{\alpha'\beta'})
\eeqar
Note that between\eqr{PairOfSUTwoKahlerForm} and 
\eqr{PairOfSLTwoKahlerForm}, $K^1$ and $K^3$ has each 
picked up a factor of $\imath$ to become real.
  From $\cal K=\cal G \cal J$ one finds the corresponding $\cal J$.  They are now
\beqar  \label {eq:PairOfSLTwoComplexStructure}
        {(\cal J^{[i]})^{\alpha,\alpha'}}_{\beta,\beta'} 
        = {((\sigma^i)^\alpha}_\beta) 
        \otimes ({\delta^{\alpha'}}_{\beta'}), \csp i = 1, 3; &\csp&
        {(\cal J^{[2]})^{\alpha,\alpha'}}_{\beta,\beta'} 
        = {((- \imath \sigma^2)^\alpha}_\beta) 
        \otimes ({\delta^{\alpha'}}_{\beta'}), \nono
        {(\cal J'^{[i]})^{\alpha,\alpha'}}_{\beta,\beta'} 
        = ({\delta^\alpha}_{\beta}) 
        \otimes {((\sigma^i)^{\alpha'}}_{\beta'}), \csp i = 1, 3; &\csp&
        {(\cal J'^{[2]})^{\alpha,\alpha'}}_{\beta,\beta'} 
        = ({\delta^\alpha}_{\beta}) 
        \otimes {((- \imath \sigma^2)^{\alpha'}}_{\beta'})
\eeqar
It is clear that they transform in the adjoint representations of $SO(1,2)$ and $SO(1,2)'$ respectively.  $O(2,2)$ elements with negative determinants exchanges them as it exchanges the two $SL(2,\RR)$'s.  Furthermore, they satisfy the following algebraic relations:
\beq    \label {eq:HomogeneousRelationOfComplexStructureJSLTwo}
         [ {\cal J}^{[i]}\ , \ \cal J'^{[j]} ] = 0 \ , 
\eeq
\beqar  \label {eq:AlgebraicRelationOfComplexStructureJSLTwo}
        {\cal J}^{[1]} {\cal J}^{[2]} = - {\cal J}^{[2]} {\cal J}^{[1]} &=&  {\cal J}^{[3]}\ , \nono
        {\cal J}^{[2]} {\cal J}^{[3]} = - {\cal J}^{[3]} {\cal J}^{[2]} &=&  {\cal J}^{[1]}\ , \nono
        {\cal J}^{[1]} {\cal J}^{[3]} = - {\cal J}^{[3]} {\cal J}^{[1]} &=&  {\cal J}^{[2]}\ , \nono
        {\cal J}^{[1]} {\cal J}^{[1]}  
                = {\cal J}^{[3]} {\cal J}^{[3]} 
                = - {\cal J}^{[2]} {\cal J}^{[2]} &=&\id \ , 
\eeqar
and the same for $\cal J'^{[i]}$.
Therefore they themselves are generators of $so(1,2)\times so(1,2)$ in the $(\bf{2},\bf{2})$ representation. This makes sense in comparison with the Euclidean case (\secr {HermitianStructureEuclidean}) because as we have just seen, $\cal J^{[1]}$ and $\cal J^{[3]}$ each picks up a factor of $\imath$.  The nine anti-commutators $\{\cal{J}^{[i]}, \cal{J}'^{[j]}\}$ do not vanish.  

An arbitrary $4\times 4$ antisymmetric matrix 
$\cal K = \cal G \cal J = \cal G \sum_i (a_i \cal J^{[i]} + a'_i \cal J'^{[i]})$.  
The condition $\cal J \cal J = - \id$ dictates that
\beq
        - (a_1)^2 + (a_2)^2 - (a_3)^2  - (a'_1)^2 + (a'_2)^2 - (a'_3)^2 = 1, \csp
        a_i a'_j = 0 \ ,
\eeq
so the only solutions are either 
\beq    \label {eq:JOnHyperbolicPlane}
        \cal J = \sum_i a_i \cal J^{[i]}, 
        \csp (a_2)^2 - (a_1)^2  - (a_3)^2 = 1 
\eeq
or
\beq
        \cal J = \sum_i a'_i {\cal J'}^{[i]}, 
        \csp (a'_2)^2 - (a'_1)^2  - (a'_3)^2 = 1 \ .
\eeq
The solutions to \eqr {JOnHyperbolicPlane} lie on a hyperboloid of two sheets, also known as pseudo-2-sphere of index 1, $S^2_1$.  We have thus found that the possible complex structures are parameterized by two hyperboloids of two sheets.  $O(2,2)$ elements with negative determinants exchange the two hyperboloids.  The two sheets of such a hyperboloid are \emph {simultaneously} mapped into each other  by an improper $SO(2,2)$ rotations.  For example, $U = U' = \sigma^1$ does that.   Note also that if a particular solution $\cal J$ lives on one sheet,  then $-\cal J$ lives on the other.

The discussion here is sufficient for flat 4d space: $\RR^4$ and its
toroidal compactifications.  For a curved space, there is the question
of the integrability of the complex structure.  The above
classification of hermitian almost complex structure applies to each
point on the manifold.

\subsubsection {Stabilizer}
\label {sec:StabilizerTwoTwo}
In this course we shall often have to consider subgroup of the
orthogonal group leaving invariant some/several subspaces of $\RR^4$.

\subsubsubsection {Stabilizer of $V^0$}
Clearly, an $O(1,2)$ subgroup of the $O(2,2)$ symmetry leaves $V^0$
invariant.  It is equally clear that an diagonal subgroup of
$SL(2,\RR)\times SL(2,\RR)$,
\beq
        U U' = \id
\eeq
leaves $V^0$ invariant in \eqr {GroupActionOnBispinorTwoTwo}.  This accounts for $SO_0(1,2)$, the proper Lorentz group.  $SO(1,2)$ is generated when one includes the operation mentioned above that reverses the signs of $V^1$ and $V^3$.  The full $O(1,2)$ is obtained by considering an parity operation such as $V^i \to - V^i$.  It corresponds to 
$\cal V_{\alpha\beta} \to - \cal V_{\beta\alpha}$.

\subsubsubsection {Stabilizer of $V^0$ and $V^2$}
The obvious $O(2)$ subgroup leaving $V^0$ and $V^2$ invariant is, in the bi-spinor form, generated by an $U(1)$
\beq
        e^{\imath \theta \sigma^2}
\eeq
and a reflection on, say, $V^3$.  

\subsubsubsection {Stabilizer of $V^0$ and $V^3$}
The obvious $O(1,1)$ subgroup leaving $V^0$ and $V^3$ invariant is in 
bi-spinor form generated by a ``boost''
\beq
        e^{\gamma \sigma^3}
\eeq
plus a reflection on, say, $V^2$.

\section {Worldsheet SCA's}

$N=2$ string theory
is obtained by gauging $N=2$ supersymmetry in two dimensions, i.e.
a two-dimensional $N=2$ supergravity \cite{Brink:1977vg}. 
Two-dimensional $N=2$ supergravity consists of one spin $2$ graviton 
field, two spin $3/2$
gravitino fields and one spin $1$ Abelian gauge field.
When one goes to the conformal gauge and uses the super-diffeomorphic
as well as super-Maxwell symmetries to remove all the degrees of freedom of the 
gauge field, graviton and gravitinos, 
one is left with $N=2$ superconformal symmetry on the worldsheet as the residual gauge symmetry.  
Such worldsheet superconformal theories  are distinguished by
the superconformal algebra (SCA) of decoupled left and right chiral currents.  
For $N=2$ string, the minimal algebra is the $N=2$ SCA.  
It 
is generated by a stress-energy tensor $T$,
two supercurrents $G^{\pm}$ and a $U(1)$ current $J$.

We will consider critical $N=2$ string theories.
Fixing the reparametrization invariance requires 
a $(b,c)$ fermionic ghost system
with $(2,-1)$ spins, 
two pairs  $(\beta^{\pm},\gamma^{\pm})$ of bosonic ghost system
with $(3/2,-1/2)$ spins, while fixing the $U(1)$ gauge symmetry requires a
 $(\tilde{b},\tilde{c})$ fermionic ghost system with $(1,0)$ spins.
The total ghost central charge is $c_{gh}=-6$. 
Consider now the matter sector. The simplest matter system arises
when the $N=2$ string propagates in a flat background. 
It is given by an $N=2$ supersymmetric $\sigma$-model
with four bosons $X^{I}$ and four fermions $\Psi^I$.
Thus, the target space is
four-dimensional. The two-dimensional $N=2$ supersymmetry
implies that the target space has a complex structure and therefore 
its signature is either $(4,0)$ or $(2,2)$. 
Only in the $(2,2)$ signature case the first quantized
string has propagating on-shell degrees of freedom and
a non-trivial dynamics in $\RR^4$.  
The Euclidean signature is nonetheless important as a simple reference 
for intuition and analysis, for curved Euclidean spaces where there are 
highly nontrivial geometric degree of freedom, and for 
its possible off-shell dynamics.
We will consider the algebraic structures of both signatures.

In the section we will review $N=2,4$ SCA in two
dimensions and consider a bigger structure with 16 supercharges.  We
shall also discuss the automorphism of these SCA's, as they will be
needed in this paper.

\subsection {$N=2$ SCA}

As noted above,
the $N=2$ supersymmetry algebra is generated by a stress-energy tensor $T$,
two supercurrents $G^{\pm}$ and a U(1) current $J$.
The singular parts of their OPEs characterize their symmetry algebra:
\bea    \label {eq:NTwoOPE}
        G^+(z)&G^+(w) &\sim~\, G^-(z)G^-(w) \,~ \sim \,~ 0 \ ,\nonumber\\
        G^+(z)&G^-(w) &\sim~\, \frac {c/3} {(z-w)^3}
        + \frac {J(w)} {(z-w)^2} + \frac {\pa J(w) /2} {z-w} +
          \frac {T(w)} {z-w} \ ,\nonumber\\
        J(z)&G^{\pm}(w) &\sim~\, \pm \frac{G^{\pm}(w)}{z-w} \ ,\nonumber\\
        J(z)&J(w) &\sim~\, \frac {c/3} {(z-w)^2} \ ,\nonumber\\
        T(z)&G^\pm(w) &\sim~\, \frac {\frac{3}{2} G^\pm(w)} {(z-w)^2}
          + \frac {\pa G^\pm(w)} {z-w} \ ,\nonumber\\
        T(z)&J(w) &\sim~\, \frac {J(w)} {(z-w)^2}
          + \frac {\pa J(w)} {z-w} \ ,\nonumber\\
        T(z)&T(w) &\sim~\, \frac {c/2} {(z-w)^4} + \frac {2 T(w)} {(z-w)^2}
          + \frac {\pa  T(w)} {z-w} \ .
\label{N2}
\eea
$c$ is the central charge normalized for the $N=0$ Virasoro algebra,
i.e. for the $\sigma$-model on $\RR^4$, $c=6$.

In term of mode operators, the above OPEs give the following $\ZZ_2$ graded Lie algebra:
\beqar    \label {NTwoSCA}
[L_m, L_n] &=& (m-n)L_{m+n} + \frac c {12} m(m^2-1) \delta_{m+n}; \nono
[L_m, J_n] &=& -n J_{m+n}; \nono
[J_m, J_n] &=& \frac c 3 m \delta_{m+n}; \nono
[L_m, G^{\pm}_n] &=& (\half m - n)G^{\pm}_{m+n}; \nono
[J_m, G^{\pm}_n] &=& \pm G^{\pm}_{m+n}; \nono
\{G^+_m, G^-_n \} &=& L_{m+n} + \frac {m-n} 2 J_{m+n} + \frac c 6
 (m^2-1) \delta_{m+n}; \nono
\{G^+_m, G^+_n\} &=& \{G^-_m, G^-_n\} ~=~ 0 \ .
\eeqar

As given above, the Abelian $R$ symmetry of the $N=2$ SCA generate by $J_0$
can be either a compact $U(1)$ or noncompact translation or ``boost.'' 
Even though the two share the same complexification, 
they are very different symmetries.  This
is reflected in the reality/Hermiticity condition imposed on the currents.
 Different conditions leads to different Hilbert spaces and hence
different theories.  In this work, as in most work on $N=2$ strings, we
consider the case of a compact $U(1)$ symmetry, so that
\beq
        (G^+_m)^\dagger = G^-_{-m} \ ,
        (J_m)^\dagger = J_{-m} \ .
\eeq
Noncompact Abelian $R$ symmetry can be studied in exactly the same vein
as the analysis done in this paper.

Besides the $U(1)$ inner automorphism generated by $J$, this algebra has
a $\ZZ_2$ automorphism
 flipping the sign of $J$ and interchanging $G^+$
with $G^-$.  
\beq    \label {eq:ConjugationAutomorphismNTwoSCA}
        T \to T, \csp J \to -J, \csp, G^\pm \to G^\mp \ .
\eeq
We call it the ``conjugation'' automorphism.  
In addition one can flip the
sign of all the odd elements of the algebra all at once.  The latter is
present for any $\ZZ_2$ graded Lie algebra.  
There are other outer automorphisms.  
One family of outer automorphisms will be used in constructing D-brane 
boundary conditions in \cite{Gluck:2003pa}.

The $N=2$ SCA has a free field realization using $d$ pairs of complex
bosons $X^i$ and fermions $\Psi^i$, $i=1\ldots d$.  
The central charge $c = 3 d$.
The only nonvanishing OPE's are
\beqar  \label {eq:NTwoFreeFieldAlgebra}
\pa X^i(z)&\pa X^{\bar j}(w) &\sim ~\frac {- \cal G^{i\bar j}} {(z-w)^2}; \nono
\Psi^i(z)&\Psi^{\bar j}(w) &\sim ~\frac {\cal G^{i\bar j}} {z-w} \ .
\eeqar

The $N=2$ generators read 
\beqar  \label {eq:NTwoFreeField}
T &=& - \cal G_{i \bar j} ~\nod {\pa X^i \pa X^{\bar j}} 
        + \frac {\cal G_{i \bar j}} 2 ~(\nod {\pa \Psi^i \Psi^{\bar j} 
        + \pa \Psi^{\bar j} \Psi^i}), \nono
J &=& \cal G_{i\bar j} \nod {\Psi^i \Psi^{\bar  j}}, \nono
G^+ &=& \imath \cal G_{i \bar j} ~\Psi^i \pa X^{\bar j}, \nono
G^- &=& \imath \cal G_{i \bar j} ~\Psi^{\bar j} \pa X^{i} \ . 
\eeqar

Even though this works exactly as it is only when
$\cal G_{i\bar j}$ is constant, $N=2$ SCA can be realized in a $N=2$
$\sigma$-model on any (pseudo-)K\"{a}hler manifold whose metric is
Ricci-flat (plus loop corrections).  \Eqr{NTwoFreeField}
provides a useful semiclassical approximation.
Since the smallness expansion parameter for such approximation is 
\beq
        \frac {\cal G_{IJ}} {\alpha'} \ ,
\eeq
one can roughly say that it is valid where the radius of spacetime is
small compared to the string length.  This is the regime where
stringiness is small and classical differential geometry of 
point-set topology prevails.

When the theory is free (except orbifolds),  i.e. when there is no
curvature in the metric or the anti-symmetric tensor $B$ field, the
conjugation automorphism can be implemented by taking $\cal G_{i\bar j}$ to
a diagonal form and 
\beq    \label {eq:FreeFieldConjugation}
        \pa X^i \leftrightarrow \pa X^{\bar i}, \csp 
        \Psi^i \leftrightarrow \Psi^{\bar j} \ .
\eeq 
This is a symmetry of the theory in target spaces with a discrete 
isometry that acts as complex conjugation on complex coordinates,
 provided one
does the same for the left and right movers.  This last restriction is due to
$X$.  $\Psi$ and $\tilde \Psi$ are independent, but $X$ cannot be
split into completely decoupled chiral and anti-chiral parts unless it
is compactified: the left and right momenta are one and the same
otherwise.  Nonetheless, it is an automorphism of the algebra.

The relation and distinction between the
symmetry of the theory and the automorphism of the chiral algebra plays
an important role in discussing the family of $N=2$ string theory in
section 3.  In this section we shall remark on such distinction when
appropriate.  As an example, note that when the target spacetime has
nontrivial holonomy, \eqr{FreeFieldConjugation}
 ceases to be a symmetry
of the algebra, because the holomorphic and anti-holomorphic tangent
bundle generically are different.  However, the $N=2$ SCA itself still
has a conjugation automorphism, and this leads to the mirror symmetry of
Calabi-Yau manifold.

We note here that there is nothing specific in the $N=2$ algebra that
requires the metric of the target space to be Euclidean in an
$N=2$ $\sigma$-model realization.  $\cal G_{i\bar j}$ can
define a non-degenerate and non-definite hermitian metric.  The only
implication of the $N=2$ structure on the signature of the metric is
that the number of temporal and spatial dimensions must both be even.

\subsection {$N=4$}
\label {sec:NIsFour}

When an $N=2$ SCFT is realized as a $\sigma$-model on target
space of complex dimension 2, the SCA is automatically extended to 
an $N=4$ SCA\footnote 
        {The $N=4$ SCA relevant to this work is what is called
        ``small'' $N=4$.}.
This of course corresponds to a vacuum
solution of the $N=2$ string, so we have to consider it.
A typical example of the target space would be a
hyper-K\"{a}hler four-manifold.
The $N=4$ SCA in fact comes in two flavors.  Realized in $\sigma$-model,
one corresponds to target space with $(4,0)$ signature, such as $K3$. 
The other corresponds to $(2,2)$ signature.  We now consider them in
turn.

\subsubsection {Compact $N=4$}
\label {sec:CompactNFour}
The $N=4$ algebra \cite{Ademollo:1976wv} 
can be obtained by supplementing an $N=2$
algebra with two additional affine currents of charges $\pm2$ with respect to
$J$.  We denote them by $J^{+}$ and $J^{-}$.
The OPE's among themselves are
\beqar  \label {eq:JPlusJMinusOPESUTwo}
J(z)~J^\pm(w) &\sim& \frac {\pm 2 J^\pm} {z-w}  \ ,\nono
 J^{+}(z) ~J^{-}(w) &\sim& \frac  {d/2} {(z-w)^2} + \frac{J(w)}{z-w} \ , \nono
J^+(z)~J^+(z) &\sim& 0 \, \nono 
J^-(z)~J^-(w) &\sim& 0\ ,
\eeqar
They obey the reality condition
\beq    \label {eq:NFourAffineCurrentReality}
        (J^+_m)^\dagger = J^-_{-m}, \csp J_m^\dagger = J_{-m}.
\eeq
Together with $J$ they 
form the affine $SU(2)$ algebra at affine level $\frac d 4$.
Defining three Hermitian chiral currents $J^i$
\beq
        J^3 = \half J; \csp J^1 \pm \imath J^2 = J^{\pm} \ ,
\eeq
the three currents will have OPE
\beq \label{eq:AffineSUTwoOPE}
        J^i(z) ~ J^j(w) \sim  \frac {\frac d 4 \delta^{ij}} {(z-w)^2} 
        + \frac {\imath \epsilon^{ijk} J^k} {z-w} \ .
\eeq

The supercharges must form a representation of this $SU(2)$.  Having the
$N=2$ SCA embedded consistently as a sub-chiral algebra means that there are four supercharges
transforming as a complex doublet of this $SU(2)$.
One can write them as $G^{\alpha,\beta}$, where $\alpha$ and $\beta$
both range over 1 to 2.  There is a
reality condition
\beq    \label {eq:RealityConditionSUTwo}
        (G^{\alpha,\beta})^\dagger
        = - {(\sigma^2)^\alpha}_{\alpha'} {(\sigma^2)^\beta}_{\beta'}
        G^{\alpha',\beta'} .
\eeq
It can be derived from the free field representation discussed below
using the bi-spinor notation discussed in section 2.

$G^{\alpha,\beta}$ have the following OPE with the $SU(2)$ currents:
\beq
        J^i(z) G^{\alpha,\beta}(w) 
                \sim  \frac {\half (\sigma^i)^\beta_\gamma 
                                G^{\alpha,\gamma}}
                {z-w} \ .
\eeq

The OPE among the $G's$ themselves is
\beq    \label {eq:NFourGGOPESUTwo}
        G^{\alpha,\beta}(z) ~ G^{\gamma,\lambda}(w) 
        \sim \frac {d \epsilon^{\alpha\gamma} \epsilon^{\beta\lambda}}
        {(z-w)^3} 
        - \frac {2 \epsilon^{\alpha\gamma} (\sigma^i)^{\beta,\lambda} 
J^i} {(z-w)^2}
        - \frac {\epsilon^{\alpha\gamma} (\sigma^i)^{\beta,\lambda} 
\pa J^i} {z-w}
        + \frac {\epsilon^{\alpha\gamma} \epsilon^{\beta\lambda} T} {z-w} \ .
\eeq
The OPE's of $(T ~ J)$ and of $(T~G)$ are the conventional ones for
chiral primary current of conformal weight $1$ and $\frac 3 2$
respectively. This completes the small $N=4$ SCA.  

It is clear from the above that the compact $N=4$ SCA has 
two $SU(2)$ automorphisms.  One acts on the first index of $G^{\alpha,\beta}$ and we call it the
outer $SU(2)$ or $SU(2)_O$.  The other acts on the second index of
$G^{\alpha\beta}$ and we shall call it the inner $SU(2)$ or $SU(2)_I$. 
$SU(2)_I$ is an inner automorphism because it is generated by
the zero modes of $J^i$.  $SU(2)_O$ is an outer automorphism because it
is not generated by any part of the $N=4$ SCA.

We note in particular that 
\beq    \label {eq:ExampleOfNTwoInNFourCompact}
        G^{1,1}(z) ~G^{2,2}(w) \sim \frac d {(z-w)^3} 
        + \frac J {(z-w)^2} + \frac {\pa J /2} {z-w}
        + \frac T {z-w} \ .
\eeq
So one embedding of $N=2$ SCA would have  
\beq
        J = 2 J^3, \csp G^+ = G^{1,1}, \csp G^- = G^{2,2}.
\eeq
This is of course by no means unique: any $SU(2)_I
\times SU(2)_O$ transformation of \eqr{ExampleOfNTwoInNFourCompact} would
give an OPE of the same type, with $J_3$ rotated in the $(3)$
representation of $SU(2)_I$.  

\subsubsection {Noncompact $N=4$}
\label {sec:NoncompactNFour}
Just as there are a compact and a noncompact version of $N=2$ SCA, the  $N=4$ SCA
discussed above has a noncompact cousin which has an affine $SL(2,\RR)$ as
its symmetry current.  As stated before, in this work we are only
concerned with $N=2$ string theories whose gauged $N=2$ SCA is the compact one.  We
have just seen how it could be embedded simply in compact $N=4$ SCA.  As
we shall demonstrate shortly, it could just as well be embedded in a
noncompact $N=4$ SCA.  Therefore it is relevant to study the latter.  In
fact it appears on any $\sigma$-model with pseudo-hyper-Ka\"hler metric,
i.e. a metric with signature $(2,2)$ that also satisfies hyper-Ka\"hler
condition.

Compared with the compact version of $N=2$ SCA, 
the different OPEs are $(J^+ ~ J^-)$, which now becomes
\beqar  \label {eq:JPlusJMinusOPESLTwo}
 J^{+}(z) ~J^{-}(w) &\sim& \frac  {- d/2} {(z-w)^2} - \frac{J(w)}{z-w} \ , \nono
\eeqar
and $(G ~ G)$, which becomes 
\beq    \label {eq:NFourGGOPESLTwo}
        G^{\alpha,\beta}(z) ~ G^{\gamma,\lambda}(w) 
        \sim \frac {d \epsilon^{\alpha\gamma} \epsilon^{\beta\lambda}}
        {(z-w)^3} 
        - \frac {2 \epsilon^{\alpha\gamma} 
        (J^2 \sigma^2 - \imath J^3 \sigma^3 - \imath J^1 \sigma^1)^{\beta,\lambda} 
J^i} {(z-w)^2}
        - \frac {\epsilon^{\alpha\gamma} (\sigma^i)^{\beta,\lambda} 
\pa J^i} {z-w}
        + \frac {\epsilon^{\alpha\gamma} \epsilon^{\beta\lambda} T} {z-w} \ ,
\eeq
where we have defined 
\beq
        J^2 = \half J \ ,
\eeq
and $J^1$, $J^2$ by\footnote
        {This is of course just a matter of convention.  We make this
        particular choice for convenience because $\imath
        {(\sigma^2)^\alpha}_\beta$ is real.}
\beq
       \csp J^3 \pm \imath J^1 = J^{\pm} \ ,
\eeq
Comparing \eqr{NFourGGOPESLTwo} with \eqr{JPlusJMinusOPESUTwo} and \eqr {NFourGGOPESUTwo}, 
we see that to obtain the new
OPE one merely need to add a factor of $\imath$ to $J^\pm$ each.
Therefore the new OPE is consistent.  However, we still require that
\eqr{NFourAffineCurrentReality} holds.  This would make no sense if the
new $J^\pm$ is literally related to the old ones by a factor of $\imath$.
Therefore they are really different affine Lie algebra of real coefficients that share the same
complexification.  Since \eqr{JPlusJMinusOPESUTwo} defines 
affine $su(2)= so(3)$, \eqr{JPlusJMinusOPESLTwo} defines affine $so(2,1) =
sl(2,\RR)$ current algebra, which can be rewritten as 
\beq    \label {eq:AffineSLTwoOPE}
        J^i(z) ~ J^j(w) \sim \frac {\half \eta^{ij}} {(z-w)^2} 
        + \frac {\imath \varepsilon^{ij}_k J^k} {z-w} \ , \nono
\eeq
The difference from \eqr{AffineSUTwoOPE} is 
that the signature $(2,1)$ metric
$\eta^{ij}$ replaces the signature $(3,0)$ metric $\delta^{ij}$ 
and that $SO(2,1)$
structure constant $\varepsilon^{ij}_k$ replaces $SO(3)$ structure constant
$\epsilon^{ijk}$.  The convention for $\eta$ and $\epsilon$ is given in
\eqr{etaConvention} and \eqr{epsilonConventionTwoOne}. 
Since $so(2,1)=sl(2,\RR)$, this is clearly the affine $sl(2)$ current algebra.

The only other difference is 
\beq    \label {eq:RealityConditionSLTwo}
        (G^{\alpha,\beta})^\dagger
        = G^{\alpha,\beta} \ .
\eeq
That is, they are all simply Hermitian, which makes sense since they are
representations of $sl(2,\RR)$, which is real.  All these will have a
concrete realization in the free field representation given in \secr {FatFreeNIsFour}.

        It is clear that the OPEs of the noncompact $N=4$ SCA are
invariant under two $SL(2,\RR)$ automorphisms.  One acts on the first 
index of $G^{\alpha,\beta}$ and we call it $SL(2,\RR)_O$.  The other
acts on the second index as well as on $J^i$ and we call it
$SL(2,\RR)_I$.  $SL(2,\RR)_I$ is an inner automorphism because it is
generated by
the zero modes of $J^i$.  $SL(2,\RR)_O$ is an outer automorphism because it
is not generated by any part of the current algebra.

We note that 
\beq    \label {eq:ExampleOfNTwoInNFourNoncompact}
        G^+ = \frac {\epsilon_{\alpha\gamma}} 2 
                {(\id - \sigma^2)^\gamma}_\beta G^{\gamma,\beta}, 
        \csp G^- = \frac {\epsilon_{\alpha\gamma}} 2 
                {(\id + \sigma^2)^\gamma}_\beta G^{\gamma,\beta}
\eeq
generate the OPE's of the compact $N=2$ SCA along with 
\beq   \label {eq:ExampleOfNTwoInNFourNoncompactJ}
        J= 2 J^2 \ .
\eeq
This is of course by no means unique: any $SL(2,\RR)_I
\times SL(2,\RR)_O$ transformation of 
\eqr{ExampleOfNTwoInNFourNoncompact} would
give an OPE of the same kind, with $J^2$ rotated in the $(3)$
representation of $SL(2,\RR)_I$.

\subsection {The fat, free $N=4$ SCA}
\label  {sec:FatFreeNIsFour}

The simplest realization of the $N=4$ SCA is the free $\sigma$-model on
$\RR^4$, which also gives the simplest $N=2$ string vacuum.  It therefore
deserves a careful study.  Upon inspection, the most interesting feature
of this model is that it contains not one, but four 
different realizations of $N=4$
SCA all embedded in equally simple ways.  In this section we will
derive this algebraic feature in detail.  The geometry of $\RR^4$ relevant
to this work is studied in details in section 2.  We shall make frequent
uses of results derived there.

There are four real bosons and four real fermions in this theory.  We
can describe both of them as $2\times 2$ matrices using the bi-spinor
notation (eq. \ref{eq:BispinorFromVectorEuclidean},
\ref{eq:BispinorFromVectorTwoTwo}).  We will consider both $(4,0)$ and
$(2,2)$ signature.  In the bi-spinor notation, the metric has the same
form for both signatures (c.f. the
discussion after \eqr {NormSquaredIsDeterminant}).
Therefore in both cases they have the following OPEs:
\beqar  \label {eq:RFourFreeFieldOPE}
        {\pa X^{\alpha,\beta}} (z) & \pa X^{\gamma,\lambda}(w) &
        \sim ~\frac {-2\epsilon^{\alpha\gamma} \epsilon^{\beta\lambda}}
                {(z-w)^2}, \nono
        {\Psi^{\alpha,\beta}} (z) & \Psi^{\gamma,\lambda}(w) &
        \sim ~\frac {2\epsilon^{\alpha\gamma} \epsilon^{\beta\lambda}}
                {z-w} \ .
\eeqar
Thus the stress tensors for both cases take the form
\beq    \label {eq:RFourStressTensor}
        T = - \frac 1 4 \epsilon_{\alpha\gamma} \epsilon_{\beta\lambda}
        \nod {\pa X^{\alpha,\beta} \pa X^{\gamma,\lambda} }
        + \frac 1 4 \epsilon_{\alpha\gamma} \epsilon_{\beta\lambda}
        \nod {\pa \Psi^{\alpha,\beta} \Psi^{\gamma,\lambda}}. \nono
\eeq
The central charge is $6$.

The difference between the two types of signatures therefore lies in the
reality condition on the free fields, which has important consequences
on the symmetry group of the theory and automorphisms of the SCA.  
Indeed, \eqr{RFourFreeFieldOPE} and \eqr{RFourStressTensor} are both 
invariant under $SL(2,\CC)$ acting on the spinor indices of
$X^{\alpha,\beta}$ and $\Psi^{\gamma,\lambda}$ respectively and
indepentently.  That is, it has four $SL(2, \mathbb {C})$
automorphisms.  
That symmetry is reduced differently for different types of signatures.
We
now consider them in turn.

\subsubsection {Euclidean Space}
\subsubsubsection {Reality}

The reality condition is, from \eqr {RealityConditionForCalVUpUpEuclidean}
\beqar  \label {eq:RFourRealityConditionEuclidean}
        (X^{\alpha,\beta})^\dagger = 
        - {(\sigma^2)^\alpha}_{\alpha'} {(\sigma^2)^\beta}_{\beta'}
        X^{\alpha',\beta'}, \nono
        (\Psi^{\alpha, \beta})^\dagger = 
        - {(\sigma^2)^\alpha}_{\alpha'} {(\sigma^2)^\beta}_{\beta'}
        \Psi^{\alpha',\beta'} ,
\eeqar
and the same for the right chiral fermions 
$\tilde \Psi^{\theta,\gamma}$. All indices range from $1$ to $2$.
The normalization has been chosen so that the metric for $X^I$ and
$\Psi^I$ with $O(4)$ vector index is identity.  The vector index is
related to the bi-spinor index by \eqr {BispinorFromVectorEuclidean}.

\subsubsubsection {Automorphisms}

\eqr{RFourRealityConditionEuclidean} breaks each of the four 
$SL(2,\CC)$ automorphism of the free OPE down to to $SU(2)$.
We shall denote the two acting on the bosons as $SU(2)_O \times
SU(2)_{O'}$, and the two acting on the fermions as $SU(2)_I \times
SU(2)_{I'}$.  In our convention, the unprimed groups act on the first
indices and the primed groups act on the second.  
They all come from the rotational symmetry of
Euclidean $\RR^4$: $SU(2)\times SU(2)' = SPIN(4)$.  
See \secr {SPINFourToSUTwoSUTwo} for more detail. 
Since the bosons are decoupled from the fermions, they can be rotated
independently. $SU(2)_I$ and
$SU(2)_{I'}$ are both symmetries of the theory, and one can rotate the
left and right chiral fermions independently.  So they are further
doubled.  We will construct the chiral currents for them shortly.  
$SU(2)_O$ and $SU(2)_{O'}$ are symmetries of the theory but there is no
chiral currents for them.  One has to rotate the whole $X$ field. 
However, the free chiral \emph {algebra} 
\eqr{RFourFreeFieldOPE}
 and its anti-chiral counterpart do
 have the automorphism of rotating $\pa X$
and $\bar \pa X$ independently, even though it is not a symmetry of the
underlying \emph {theory}.

As discussed in \secr {ParityOperationEuclidean}, parity operations
in O(4) exchange $SU(2)$ and $SU(2)'$.  This is reflected in the field
theory as well:
\eqr{RFourFreeFieldOPE} also automorphisms that swap 
the two indices of $\pa X^{\alpha,\beta}$ and/or those of
$\Psi^{\gamma,\lambda}$.  Again, corresponding to them are a set
of symmetries for the left and right fermions $\Psi$'s respectively and
independently, but just one set for the bosons $X$'s.

\subsubsubsection {$N=4$ SCAs}

Normal-ordered bilinear of the $\Psi$'s give rise to $6$ independent
fields.  They form a pair of commuting affine $SU(2)$ currents.
\beqar  \label {eq:PairOfSUTwoCurrent}
        J^i &=& \frac 1 8 
        (\sigma^i)_{\alpha\gamma} \epsilon_{\beta\lambda} 
        \nod {\Psi^{\alpha,\beta} \Psi^{\gamma,\lambda}} 
        = \frac \imath 4 {\cal K^i}_{\alpha,\beta;\lambda,\gamma} 
                \nod {\Psi^{\alpha,\beta} \Psi^{\gamma,\lambda}}\ , \nono
        J'^i &=& \frac 1 8 (\sigma^i)_{\beta\lambda} 
        \epsilon_{\alpha\gamma} \nod {\Psi^{\alpha,\beta} \Psi^{\gamma,\lambda}}
        = \frac \imath 4 {\cal K'^i}_{\alpha,\beta;\lambda,\gamma} 
                \nod {\Psi^{\alpha,\beta} \Psi^{\gamma,\lambda}} \ .
\eeqar
$\cal K^i$ are defined in \eqr{PairOfSUTwoKahlerForm}.  
The OPEs of $J$ are 
\beqar  \label{eq:JJPrimeOPESOThree}
        J^i(z) ~ J^j(w) &\sim& \frac {\half \delta^{ij}} {(z-w)^2} 
        + \frac {\imath \epsilon^{ijk} J^k} {z-w}, \nono
        J'^i(z) ~ J'^j(w) &\sim& \frac {\half \delta^{ij}} {(z-w)^2} 
        + \frac {\imath \epsilon^{ijk} J'^k} {z-w}, \nono
        J^i(z) ~ J'^j(w) &\sim& 0 \ .
\eeqar
It is easy to see that $J^{i}$ and $J'^{i}$ are respectively the affine currents for $SU(2)_I$ and $SU(2)_{I'}$.

Since $X$ and $\Psi$ are free fields we can make sixteen supercharges
from bilinears of them.  They are conveniently grouped into four
quadruplet:
\beqar  \label {eq:AllTheSupercurrentEuclidean}
        G_{00}^{\alpha,\gamma} &=& \frac \imath 2 \epsilon_{\beta\lambda}
        \pa X^{\alpha,\beta} \Psi^{\gamma,\lambda}, \nono
        G_{10}^{\beta,\gamma} &=& \frac \imath 2 \epsilon_{\alpha\lambda}
        \pa X^{\alpha,\beta} \Psi^{\gamma,\lambda}, \nono
        G_{01}^{\alpha,\lambda} &=& \frac \imath 2 \epsilon_{\beta\gamma}
        \pa X^{\alpha,\beta} \Psi^{\gamma,\lambda}, \nono
        G_{11}^{\beta,\lambda} &=& \frac \imath 2 \epsilon_{\alpha\gamma}
        \pa X^{\alpha,\beta} \Psi^{\gamma,\lambda}.
\eeqar

The OPEs between the $G$'s from different quadruplet give rise to exotic
spin two fields but they are not used in the work.  What is relevant is
the observation that each of the four quadruplet alone is a candidate
for the group of $4$ supercurrent in the $N=4$ SCA.  For example, the
OPE's among $G_{00}$ reproduce \eqr{NFourGGOPESUTwo} with $J^{i}$
from \eqr{PairOfSUTwoCurrent} as the affine $SU(2)$ currents; 
so do those among $G_{01}$.  On the other hand, the
OPE's among $G_{10}$ or $G_{11}$ reproduce the same algebra but with
$J'^{i}$ in lieu of $J^{i}$.  
There are thus four simple ways to embed the compact $N=4$ SCA
in this big algebra.  The stress tensor is of course always the same:
\eqr{RFourStressTensor}. The affine currents and the supercurrents can
be grouped as $(J, G_{00})$, $(J, G_{01})$, $(J', G_{10})$, or $(J', G_{11})$.  
It is
clear that depending on the grouping, one of $SU(2)_I$ and $SU(2)_{I'}$
becomes the inner $SU(2)$ of the $N=4$ SCA, while one of $SU(2)_O$ and
$SU(2)_{O'}$ becomes the outer $SU(2)$.

\subsubsection {$(2,2)$}
\subsubsubsection {Reality}

The reality condition is, from \eqr {RealityConditionForCalVUpUpTwoTwo}
\beq    \label {eq:RFourRealityConditionTwoTwo}
        (X^{\alpha,\beta})^\dagger
        = X^{\alpha,\beta}, \csp
        (\Psi^{\alpha,\beta})^\dagger
        =\Psi^{\alpha,\beta} \ .
\eeq

\subsubsubsection {Automorphisms}

\eqr{RFourRealityConditionTwoTwo} breaks each of the four 
$SL(2,\CC)$ automorphism of the free OPE down to to $SL(2,\RR)$.
We shall denote the two acting on the bosons as $SL(2,\RR)_O \times
SL(2,\RR)_{O'}$, and the two acting on the fermions as $SL(2,\RR)_I \times
SL(2,\RR)_{I'}$.  In our convention, the unprimed groups act on the first
indices and the primed groups act on the second.  
They all come from the rotational symmetry of
$(2,2)$ space : $SL(2,\RR)\times SL(2,\RR)' = SPIN(2,2)$.  
See \secr {SPINTwoTwoToSLTwoSLTwo} for more detail. 
Since the bosons are decoupled from the fermions, they can be rotated
independently. $SL(2,\RR)_I$ and
$SL(2,\RR)_{I'}$ are both symmetries of the theory, and one can rotate the
left and right chiral fermions independently.  So they are further
doubled.  We will construct the chiral currents for them shortly.  
$SL(2,\RR)_O$ and $SL(2,\RR)_{O'}$ are symmetries of the theory but there is no
chiral currents for them.  One has to rotate the whole $X$ field. 
However, the free chiral \emph {algebra} 
\eqr{RFourFreeFieldOPE}
 and its anti-chiral counterpart do
 have the automorphism of rotating $\pa X$
and $\bar \pa X$ independently, even though it is not a symmetry of the
underlying {\emph theory}.

As discussed in \secr {ParityOperationTwoTwo}, parity operations
in O(2,2) exchange $SL(2,\RR)$ and $SL(2,\RR)'$.  This is reflect in the field
theory as well:
\eqr{RFourFreeFieldOPE} also admits automorphisms that swap
the two indices of $\pa X^{\alpha,\beta}$ and/or those of
$\Psi^{\gamma,\lambda}$.  Again, corresponding to them are a set
of symmetries for the left and right fermions $\Psi$'s respectively and
independently, but just one set for the bosons $X$'s.

\subsubsubsection {$N=4$ SCAs}

Normal-ordered bilinear of the $\Psi$'s give rise to $6$ independent
fields.  They form a pair of commuting affine $SL(2,\RR)$ currents.
\beqar  \label {eq:PairOfSLTwoCurrent}
        J^i 
        = \frac \imath 4 {\cal K^i}_{\alpha,\beta;\lambda,\gamma} 
                \nod {\Psi^{\alpha,\beta} \Psi^{\gamma,\lambda}}\ , \nono
        J'^i 
        = \frac \imath 4 {\cal K'^i}_{\alpha,\beta;\lambda,\gamma} 
                \nod {\Psi^{\alpha,\beta} \Psi^{\gamma,\lambda}} \ .
\eeqar
$\cal K^i$ are defined in \eqr{PairOfSLTwoKahlerForm}.  
The OPEs of $J$ are
\beqar  \label{eq:JJPrimeOPESOOneTwo}
        J^i(z) ~ J^j(w) &\sim& \frac {\half \eta^{ij}} {(z-w)^2} 
        + \frac {\imath \varepsilon^{ij}_k J^k} {z-w}, \nono
        J'^i(z) ~ J'^j(w) &\sim& \frac {\half \eta^{ij}} {(z-w)^2} 
        + \frac {\imath \varepsilon^{ij}_k J'^k} {z-w}, \nono
        J^i(z) ~ J'^j(w) &\sim& 0 \ .
\eeqar
Comparing with \eqr{JJPrimeOPESOThree} it is clear that $J^i$ and $J'^i$ are 
respectively the affine currents for $SL(2,\RR)_I$ and $SL(2,\RR)_{I'}$.

Since $X$ and $\Psi$ are free fields we can make sixteen supercharges
from bilinear of them.  They are conveniently grouped into four
quadruplet:
\beqar  \label {eq:AllTheSupercurrentTwoTwo}
        G_{00}^{\alpha,\gamma} &=& \frac \imath 2 \epsilon_{\beta\lambda}
        \pa X^{\alpha,\beta} \Psi^{\gamma,\lambda}, \nono
        G_{10}^{\beta,\gamma} &=& \frac \imath 2 \epsilon_{\alpha\lambda}
        \pa X^{\alpha,\beta} \Psi^{\gamma,\lambda}, \nono
        G_{01}^{\alpha,\lambda} &=& \frac \imath 2 \epsilon_{\beta\gamma}
        \pa X^{\alpha,\beta} \Psi^{\gamma,\lambda}, \nono
        G_{11}^{\beta,\lambda} &=& \frac \imath 2 \epsilon_{\alpha\gamma}
        \pa X^{\alpha,\beta} \Psi^{\gamma,\lambda}.
\eeqar

The OPEs between the $G$'s from different quadruplet give rise to exotic
spin two fields but they are not used in the work.  What is relevant is
the observation that each of the four quadruplet alone is a candidate
for the group of $4$ supercurrent in the $N=4$ SCA.  For example, the
OPE's among $G_{00}$ reproduce \eqr{NFourGGOPESLTwo} with $J^{i}$
from \eqr{PairOfSLTwoCurrent} as the affine $SL(2,\RR)$ currents; 
so do those among $G_{01}$.  On the other hand, the
OPE's among $G_{10}$ or $G_{11}$ reproduce the same algebra but with
$J'^{i}$ in lieu of $J^{i}$.  
There are thus four simple ways to embed the noncompact $N=4$ SCA
in this fat, free noncompact $N=4$ algebra.  
The stress tensor is of course always the same:
\eqr{RFourStressTensor}. The affine currents and the supercurrents can
be grouped as $(J, G_{00})$, $(J, G_{01})$, $(J', G_{10})$, or $(J', G_{11})$.  
It is
clear that depending on the grouping, one of $SL(2,\RR)_I$ and $SL(2,\RR)_{I'}$
becomes the inner $SL(2,\RR)$ of the $N=4$ SCA, while one of $SL(2,\RR)_O$ and
$SL(2,\RR)_{O'}$ becomes the outer $SL(2,\RR)$.

\section{$\alpha$, $\beta$, etc.}     
\label {sec:AlphaBeta}

\subsection {Principle and Recipe}
\label {sec:PrincipleAndRecipe}

In this section we will study the families of $N=2$
strings on $\RR^4$ and $\RR^{2,2}$  as well as its toroidal compactifications.  We propose an ansatz of
a simple realization of  the $N=2$ SCA and find all solutions.  They fall into
two broad disjoint classes, which we respectively name as 
$\alpha$ and $\beta$ strings. 
We will study their parameters of gauging.  Finally we will 
generalize the notion of the $\alpha$-$\beta$ divide to general spacetime
with a precise and topological definition. 

\subsubsection {Pre-theory and gauge theory}
\label {sec:Principle}
$N=2$ SCFT with a central charge of $6$ is the most usual starting point for studying $N=2$ string theory, but the two are really different theories, even from a worldsheet point of view.  To string theory, the gauged chiral algebra is the remnant of the worldsheet gauge symmetry.  Residual gauge symmetry 
represents the redundancy in the CFT description of the system not removed by the conformal gauge fixing.  The true physical content of the $N=2$ string worldsheet theory therefore resides in the quotient of the CFT by the $N=2$ SCA\footnote{ 
        The word quotient here has a precise meaning in the language
        of Hamiltonian formulation.  It is tantamount to imposing constraints
        as in bosonic string and superstring theories.}.
If there are different ways to embed $N=2$ SCA in its chiral algebra, to specify an $N=2$ string theory we have to specify which one to gauge.  Different embedding, however, does not necessarily leads to inequivalent quotients.  
If one  embedding is related to another by a symmetry of the theory, their
quotient will related by a relabeling of fields corresponding to  that symmetry and cannot represent any physical difference.  Otherwise, the resulting theories are really physically distinct.

The above consideration leads to a principle that is applicable to all string theories and in fact all theories obtained from another, which we shall call the \emph{pre-theory},  by gauging  some symmetry algebra therein:
\begin{quote}
        \emph {Pre-theory is supposed to be a redundant description of
the physical degree of freedom and dynamics. The redundancy is
expressed by the gauge symmetry.  
\[   \mbox {Physical Theory} 
    = \frac {\mbox {Pre-theory}} {\mbox{Gauge Symmetry}}
\]
If we are given a pre-theory and asked to find the physical theory,  we
have to known not only the intrinsic property of the gauge symmetry,
i.e. what it is, but also its extrinsic property as a part of the
pre-theory, i.e. how it is realized --- how it acts.  Both information
together determine what is redundant and what is not in the pre-theory.
One pre-theory and one type of gauge symmetry  
can thus lead to a family of physically
        distinct theories.  The space of such theories is given by the
        quotient of the space of all realization of the gauge symmetry in the
        pre-theory, quotiented by all the symmetry of the theory.  In other
        words, the space of physically distinct theories is the space of
        orbits by the action of all symmetries of the pre-theory on the space
        of all realizations of the gauge symmetry.  
        We call it the \underline {parameter space of gauging.}}
\end{quote}
What is special about the $N=2$ string is that it readily provides examples of  nontrivial family of physically inequivalent embedding due to both its rich structure and some field-theoretic ``coincidence.''
Although not logically necessary, the most studied pre-theories of
$N=2$ strings have space(-time) interpretation because they
have more geometric as well as physical relevance.  They are the $N=2$
$\sigma$-models with some $4d$ target space $\fX$.  The dimension $4$
is due to the relation between central charge, required to be 6 by
anomaly cancellation, and the dimension of the spacetime. It happens
that the $N=2$ superconformal invariance for such models often implies the
existence of the bigger $N=4$ SCA.  There is not just one, but a
continuous range of possible embeddings of  the $N=2$ SCA into the $N=4$. 
Furthermore, the $N=2$ string propagating in flat 4d spaces has the
"fat,  free" $N=4$ SCA, which yields even more choices for embedding an
$N=2$ SCA. In general, exhaustion of all embeddings of a particular
algebra in a bigger algebra may require a massive effort when the
algebras under question are infinite-dimensional.  However we found
that by restricting ourselves to \emph{simple} realizations or
embeddings of the $N=2$ SCA, to be defined precisely below\footnote{
        These embeddings are explicitly given in \eqr{SimpleNTwoRealization} for        the flat space, and\eqr{SimpleEmbeddingNTwoInNFourCompact}      and\eqr{SimpleEmbeddingNTwoInNFourNoncompact} for more general  cases.},
we can completely classify the space of all physically distinct $N=2$
string theories and they are nontrivial.  We may add that although we
do not claim their nonexistence, we are not yet aware of any
realizations that are not simple and hence fall out of our
classification.  In any case the simple realizations are the one with
the most direct geometric interpretation.

\subsubsection {Symmetries of a $\sigma$-model}
\label {sec:Recipe}
The first step in the classifications we undertake is to identify all possible realizations of $N=2$ SCA satisfying our assumption of  ``simplicity.''  After that we have to find out how the symmetries of the CFT acts on them so we can find the orbits of this action.   The first step is determined by the chiral algebra in which we are  to embed the $N=2$ SCA and is the more straightforward step because our simplicity condition is fairly restrictive.  The second step is subtler and here we give a detailed general discussion.

In a supersymmetric $\sigma$-model the bosonic fields $X^I$ define a map from the two-dimensional worldsheet into the target space $\fX$; while the fermions
$\Psi^I$ defines a section in the pull back of the tangent bundle of
the $\fX$ induced by the map $X$.  The symmetries of a model as
geometric as such are intimately related to the geometry of the target
space, but the very different nature of $X^I$ and $\Psi^I$ brings about
rather different relations.  Every isometry of the target space, i.e.
coordinate transformations that leaves the metric invariant,
corresponds to a symmetry acting on $X^I$.  The induced action on the
tangent bundle corresponds to an accompanying action on $\Psi$.   There
are also symmetries on the  $\Psi$ alone.  In a target space
interpretation they would be a rotation that leave the metric
invariant, i.e. an element of the orthogonal group of appropriate
signature.  Furthermore this rotation needs to leave the affine
connection on $\fX$ invariant; otherwise it would not leave the
$\sigma$-model action invariant.  So the kind of symmetry group is the
commutant of the holonomy group of $\fX$ within the appropriate
orthogonal group.  If, for example, $\fX$ is a generic K\"{a}hler
manifold, the  symmetry for the fermions would be an $U(1)$.  In fact,
since the left and right chiral fermions on the worldsheet decouple,
they can be rotated independently, the number of symmetries is
doubled.  Contrast this with $X$.  The left and right moving parts of
$X$ do not decouple.  Therefore the symmetries acting on $X^I$ is in
one to one correspondence with the isometries of $\fX$ which  gives
rise to them.\footnote{
        There are cases, for example with the target space a torus, when there are      additional symmetry for the $X$'s.  However, they do not have a direct  geometric interpretation in the target space.  For torus, a T-duality links them        with isometry of the T-dual target space.} 
$\sigma$-model on group manifold with the appropriate WZNW term is such an example.

Finally, when the worldsheet theory has parity invariance, exchanging
left and right mover is a symmetry.  For a $\sigma$-model, standard
worldsheet parity symmetry holds if and only the antisymmetric tensor
$B$ field vanishes.  If so, we should mod out the exchange symmetry of
$\cal J$ and $\tilde \cal J$ to obtain the space of gauge parameters
without duplication.  We note that sometimes a modified version of
worldsheet parity transformation exists, even though the standard parity
symmetry is broken by the presence of $B$ field.  The unbroken symmetry
is a parity transformation plus a rearrangement of say the right movers.
 This happens on compactification over a torus with $B$ field that is
T-dual to a non-rectangular torus without $B$ field.  One should take
all such symmetry, indeed all symmetries of the CFT into account when
determining  the parameter space of gauging for a specific case, although the preceding list seems to cover all known cases.

These considerations lead to the following recipe for finding different $N=2$ string theories.
\begin  {quote} \emph{
\noindent  Find all  realizations of the $N=2$ SCA in the pre-theory.
Quotient out by the symmetries only acting on the fermions, which is
the commutant of the holonomy group of the target space.  This can be
done independently for the left and right movers 
and gives the same reduced space of embedding for each.  
The total reduced space is their product.  Next quotient out by the 
isometries of the target space $\fX$.  They often act on the
left and right chiral algebra simultaneously and therefore
quotienting by it may intertwines the two reduced spaces of
embeddings.  Finally, if the theory has some symmetry involving
worldsheet parity we should mod it out as well.  This will exchange
the left and right movers' reduced spaces of gauging parameters up to 
some transformation.}
\end {quote}

As we consider different target space geometries, the number of symmetries affecting the $X$ and $\Psi$ vary from case to case.  This affects the range of distinct $N=2$ string vacua in two ways. On the one hand the number of ways to embed an $N=2$ SCA in the chiral algebra changes.  Generally it decrease when the theory becomes less symmetric.  On the other hand the number of symmetries that relates these different embeddings also decrease.  In this section we will analyze the case of flat space thoroughly.  It is not only the simplest but perhaps also the most important physically.  

\subsection {Simple Realizations of the $N=2$ SCA}
\label {sec:NTwoInRFourSimpleRealization}
As a first step we want to find the space of different embeddings of 
the $N=2$ SCA in the free $\sigma$-model on $\RR^4$.  Because the
infinite-dimensional nature of the operator algebra a \emph{rigorous}
and \emph{exhaustive} survey of all possible embeddings of the $N=2$
SCA  in a given theory is technically rather difficult, even for a free
theory.  So here we shall instead prove a result for an embedding of
the $N=2$ SCA satisfying certain simplifying assumptions. 
Fortunately,  this turns out to give nontrivial examples already.  
Other cases, if they exist, would be more subtle.  The analysis takes
advantage of the fact that the theory is free.  When one is dealing
with an interacting CFT, such as
in curved spacetime \secr{CurvedSpace}, one has to rely on the abstract knowledge of the symmetry currents.  The methodology will be developed therein.

        As we reviewed in section 2, the $N=2$ SCA has a free field realization \eqr{NTwoFreeField}.  Generalizing this we shall consider realizations of $N=2$ SCA currents in terms of normal ordered bilinears of free fields.
\beqar  \label {eq:SimpleNTwoRealization}
T &=& - \half \cal G_{I J} ~\nod {\pa X^I \pa X^{J}} 
        + \frac {\cal G_{I J}} 2 ~(\nod {\pa \Psi^I \Psi^{J} 
        + \pa \Psi^{J} \Psi^I}), \nono
J &=& \frac {\imath} 2 {\cal K}_{IJ}\nod {\Psi^I \Psi^{J}}, \nono
G^+ &=& \imath m_{I J} ~\Psi^I \pa X^{J}, \nono
G^- &=& \imath m^*_{I J} ~\Psi^{I} \pa X^{J} \ . \nono
\eeqar
We call such free field realization ``simple.'' The (conjugation) relation between $G^\pm$ is required by the compactness of the $U(1)$ symmetry associated with $J$.

Let $m_{IJ}$ be an arbitrary $D\times D$ complex matrix.  $D=2d$ is the real dimensions of the target space.  The OPE of $(G^+~G^+)$ and $(G^-~G^-)$ should vanish.  This leads to the following conditions on $m$:
\beq
        m \cal G^{-1} m^\top = m^\top \cal G^{-1} m= 0.
\eeq
OPE of $(G^+~G^-)$ is given by \eqr{NTwoOPE}, so we find 
\beq
        m \cal G^{-1} m^\dagger + m^* \cal G^{-1} m^\top
        = m^\top \cal G^{-1} m^* + m^\dagger \cal G^{-1} m = \cal G,
\eeq
and 
\beq
        \imath \cal K = (m \cal G^{-1} m^\dagger - m^* \cal G^{-1} m^\top).
\eeq
The OPE of $(J~G^\pm)$ results in the conditions
\beq
   \imath\cal K \cal G^{-1} m = m, \csp\csp \imath\cal K \cal G^{-1}m^* =-m^*.
\eeq
and the OPE of $(J~J)$ gives 
\beq
        \tr (\cal K g^{-1})^2 = D.
\eeq

Define ${\cal M}=\cal G^{-1}m$.  The above condition on $m$ can be rewritten as
\beqar  \label {eq:MgM}
        {\cal M} \cal G^{-1} {\cal M}^\top \,
        = \,{\cal M}^\top \cal G^{-1} {\cal M} &=& 0, \nono
        {\cal M} \cal G^{-1} {\cal M}^\dagger 
        + {\cal M}^* \cal G^{-1} {\cal M}^\top &=& \cal G^{-1}, \nono
        {\cal M}^\top \cal G {\cal M}^* 
        + {\cal M}^\dagger \cal G {\cal M} &=& \cal G.
\eeqar
and 
\beq    \label {eq:KMg}
        \imath \cal K {\cal M} = \cal G {\cal M}, \csp \imath \cal K {\cal M}^* = - \cal G {\cal M}^*,
\eeq
with
\beq    \label {eq:gkg}
        \imath \cal G^{-1} \cal K \cal G^{-1}
        = {\cal M} \cal G^{-1} {\cal M}^\dagger 
        - {\cal M}^* \cal G^{-1} {\cal M}^\top.
\eeq

One then finds that 
\beq
        ({\cal M}+{\cal M}^*)\cal G^{-1} ({\cal M}+{\cal M}^*)^\top = \cal G^{-1}
\eeq
Therefore ${\cal M}+{\cal M}^*$ leaves the metric invariant:
it must be an element of $O(4,0)$ or $O(2,2)$.  Since 
\[      G^+ = \imath {\cal M^I}_{ J} ~\Psi_I \pa X^{J} \ ,  \csp 
        G^- = \imath {{\cal M^{*I}}_{ J}} ~\Psi_I \pa X^{J} \]
we can assume without loss of generality that 
${\cal M}+{\cal M}^* = \id$ by rotating  $\Psi^J$, \footnote
        {Note since the left and right chiral fermions can be rotated
        independently, this assumption can be made for both movers
        independently and simultaneously.}
Doing so eliminates some of the symmetries that relate different embeddings.
Therefore we can write 
\beq
        {\cal M} = \half (\id + \imath {\cJ})
\eeq
for some real matrix $\cJ$, i.e. 
\beq
        {\cJ} = -\imath ({\cal M} - {\cal M}^*).
\eeq
One infers from \eqr{gkg} that 
\beq    \label {eq:kInTermsOfgJ}
        \cal K = \half (\cal G J - J^\top \cal G) \ .
\eeq
It then follows that ${\cJ}$ satisfy
\beq
        {\cJ}^\top \cal G {\cJ} = \cal G, \csp \cJ \cJ = -\id.
\eeq
Thus ${\cJ}$ is a complex structure hermitian with respect to the
metric $\cal G$.  Furthermore it follows either from \eqr{MgM} or from 
\eqr{kInTermsOfgJ} and \eqr{KMg} that
\beq    \label {eq:KahlerForm}
        \cal K = \cal G {\cJ}
\eeq
is the K\"{a}hler form.  All the other equations are satisfied.

Therefore we want to find all possible complex structures
$\cal J$ that makes $\cal G$ hermitian. 
This mathematical problem has already been solved in \secr 
{HermitianStructureEuclidean} and \secr {HermitianStructureTwoTwo}. 
For $(4,0)$ metric the solutions lies on two disjoint $S^2$ while for $(2,2)$ metric the solution lies on two disjoint $S^2_1$.  \Eqr{SimpleNTwoRealization} and \eqr{KahlerForm} then determine the affine $U(1)$ current in terms of the choice for the complex structures.  Moreover upon comparing \eqr{PairOfSUTwoCurrent} with \eqr{PairOfSUTwoKahlerForm} and \eqr {PairOfSUTwoCurrent} with\eqr{PairOfSUTwoKahlerForm} one sees that the pair of affine $SU(2)$ ($SL(2,\RR)$) currents is directly to the $SU(2)$ ($SL(2,\RR)$) 
algebra complex structures Hermitian with respect to a $(4,0)$ ($(2,2)$)
metric.  To make an $N=2$ SCA the affine $U(1)$ has to come from either
of the pair, but not a linear combination of both.  Note also that $\cal J$ and $-\cJ$ 
defines the same $N=2$ SCA: it corresponds to the
conjugation automorphism of \eqr{ConjugationAutomorphismNTwoSCA}.  
Quotiented by this $\ZZ_2$, $S^2$ becomes the real projective plane
$RP^2$, and $S^2_1$ becomes the upper sheet of its former self,
$S^2_{1+}$. Therefore the space of choices is actually reduced by
half to two disjoint $RP^2$ for $(4,0)$ metrics, and two disjoint
$S^2_{1+}$ for $(2,2)$ metrics.  
We denote them by $RP^2 \union RP^2{}'$ and 
$S^2_{1+} \union S_{1+}^2{}'$ respectively, 
where the meaning of ' is hopefully clear.  
Note also that this space is already reduced from the the total space
of different embeddings because we have used the symmetry of the theory
that rotate $\Psi^I$ to make $\cal M + {\cal M}^* = \id$.

Since the worldsheet theory includes both the left and right movers, 
each side has this same amount of choices, 
and the total space of choices is 
the square of what we have just found.  So all
in all, we have found that partially reduced parameter space of gauging
for Euclidean flat spaces to be
\beq    \label {eq:PartialReducedFlatSpaceChoicesEuclidean}
    (RP^2 \union RP^2{}') \times (RP^2 \union RP^2{}') 
    =\left(RP^2 \times RP^2{}' \union RP^2{}' \times RP^2\right) \union 
     \left(RP^2 \times RP^2 \union RP^2{}' \times RP^2{}'\right)\ .
\eeq
For flat $(2,2)$ spaces it is 
\beq    \label {eq:PartialReducedFlatSpaceChoicesTwoTwo}
    (S^2_{1+}\union S_{1+}^2{}') \times (S^2_{1+}\union S_{1+}^2{}') = 
    \left(S^2_{1+} \times S^2_{1+}{}' \union S^2_{1+}{}' \times S^2_{1+}\right) \union 
     \left(S^2_{1+} \times S^2_{1+} \union S^2_{1+}{}' \times S^2_{1+}{}'\right).
\eeq
Already we can define what we mean by  $\alpha$ and $\beta$ strings by
looking at the RHS.   $\beta$ strings choose 
$\cal J$ and $\tilde\cal J$ from the same $S^2$ or $S^2_{1+}$, 
while $\alpha$ strings choose from different ones.  
Hence they respectively belong to the sets  grouped in the first and
second pairs of parentheses in the RHS of the equations above.
  From this we now have to find the orbits of the symmetries 
acting on $X^I$, 
Certainly this does not only depend on what the symmetry group is 
but also how it acts on the reduced parameter space.
We will start analyzing case by case  noncompact and compact
flat spaces. One remark is warranted here.   As mentioned in the recipe
given in \secr{Recipe}, if the worldsheet theory  has some symmetry
involving worldsheet parity change,  the left and right chiral parts
can be interchanged.  We should then quotient the product above by it. 
The only term in this free field theory that can violate parity is  the
presence of a nonvanishing but flat $B$ field.  It has no effect on 
the OPEs and therefore has not entered into the discussion until now.  
However, even with a nonvanishing $B$ field it is possible to have a 
symmetry that involves the flip of worldsheet parity \emph{and}  a
rotation of, say, the right moving fields, while each alone is not a
symmetry of the worldsheet theory.  We will discuss this important
point further  when we discuss toroidal compactification of two or more
directions in \secr{ToroidalCompactification} because only in these
situations can there be background for $B \neq 0$ that is not a gauge 
artifact.

\subsection {Choice mod Symmetry: $\alpha$ versus $\beta$}
\label {sec:RFour}
For $\RR^4$, the rotational isometry to quotient is $O(4)$ or $O(2,2)$, 
depending on the signature of the metric.  We study them below 
separately.

\subsubsection {Euclidean}
Since we have already fixed $M+M^*$ to be $\id$, the symmetries 
that are left are the rotations that leave this invariant.  As 
\beq
        \pa X^I \to {\Lambda^I}_J \pa X^J, \csp 
        \Psi^I \to {\Gamma^I}_J \Psi^J, \csp
        \cal {M^I}_J = {(\id + \imath J)^I}_J \to {(\Gamma M \Lambda)^I}_J ,
\eeq
what is allowed is a diagonal action of the rotations on $\Psi$ and $X$: \beq
        \Gamma \Lambda = \id \ .  
\eeq
The need to involve rotations on $X$ means that one cannot independently
perform the allowed rotations separately for left and right movers.
Included in the rotations are parity operations that swap $\cJ$ and $\cJ'$, as
well as arbitrary $SO(4)$ action.  Therefore we may fix 
$\cal J$ to be, say, $\cal J^{[3]}$.  This choice is still invariant under $SU(2)'$ and a $U(1)$ subgroup of $SU(2)$ generated by $J^3$.  

Now look at the right movers.  We no longer have the freedom to pick 
which sphere of complex structure to use for $\tilde \cal J$,  
because doing so requires
swapping the two $SU(2)$ indices of  $X^{\alpha,\beta}$ (\secr{ParityOperationEuclidean}), but this would upset the choice already made for the left mover.  Therefore just as argued in the section 3.1, 
\begin {quote} \emph{there are two
broad types of $N=2$ strings in $\RR^4$.  In one the left and right
chiral $U(1)$ currents respectively use complex structures on two 
\underline {different} spheres, and we call it the $\alpha$-string;  
in the other they are from the \underline {same} sphere, and we call it the  
$\beta$-string\footnote{
        It should be obvious that what matters is how one complex
        structure differ relative to the other.  A parity symmetry
        operation on $X$ cause both sides to hop to the other sphere.}.}
\end {quote}
This definition is compatible with the earlier grouping of the 
components of the partially reduced parameter space of gauging.

For $\alpha$-string, we can rotate $\tilde \cJ$ to a particular point of 
${RP^2}'$ by an $SU(2)'$ rotation, which does not affect the already 
made choice for $\cal J$.  So there is no more physical inequivalent 
choices to be made.

For $\beta$-string, it is impossible to fix $\tilde \cal J$ completely 
because the only rotation left is the $U(1)$ that rotates $\tilde\cJ$
around $\cJ$.  It leaves $\cJ$ invariant. Since $\cJ$ and $\tilde \cJ$ use the
same sphere of complex structures it makes sense to compare them, i.e.
measure their difference by the angle between them.  In fact this
parameterizes the orbits of $\tilde \cJ$ under the residual $U(1)$.
Therefore the choices is represented by a variable $\theta$ ranging from
$0$ to $\pi$, measuring the angle between the left and right moving
$\cJ$ on the sphere.  The range of $\theta$ is further reduced to $0
\ldots \frac \pi 2$ because, as mentioned above, inequivalent choices
lie on $RP^2 = S^2/Z_2$ instead of $S^2$.  We call the $N=2$ strings
theories with this choice as type $\beta_\theta$.

Starting with the partially reduced parameter space in
\eqr{PartialReducedFlatSpaceChoicesEuclidean}, what we have just shown
is in algebraic terms
\beq
    \frac {(RP^2 \union RP^2{}') \times (RP^2 \union RP^2{}')}
    {\ZZ_2 \times O(4)}
    = \{\mbox{1 point} \} \union 
    \frac {RP^2 \times {RP^2}} {SU(2)}
\eeq
On the RHS first part of
the union corresponds to the $\alpha$ string;  the second to the $\beta$
string. The $\ZZ_2$ on the LHS is the worldsheet parity which 
exchanges $\cal J$ and $\tilde\cJ$.  There is no relevant $B$ 
field background in empty $R^4$ and so it is a symmetry. 
Without it the RHS would have been doubled.

\subsubsection {$(2,2)$}
Much of the analysis for $(4,0)$ signature carries over to $(2,2)$
signature, with $SL(2, \RR)$ replacing $SU(2)$, so we shall be brief and
emphasize only the differences.  By diagonal $O(4,4)$ action on $X$ and
$\Psi$, we can again fix
\beq
        \cal J = \cal J^{[2]}.
\eeq
For the right movers, there is again a discrete choice between putting
$\tilde\cJ$ on the same hyperboloids or the other one.  $\beta$-string chooses
the same and $\alpha$-string chooses different.

For $\alpha$-string, the residual $SL(2,R)'$ can fix $\tilde\cJ$ once one decides on which sheet of the second hyperboloid it should lie.  However, the choice between
the two sheets of the same hyperboloid is not really a choice for
different embeddings of $N=2$ SCA.  It corresponds to the $\ZZ_2$
conjugation automorphism of $N=2$ SCA
\eqr{ConjugationAutomorphismNTwoSCA} and hence gives the same $N=2$ SCA.
 Therefore the physically distinct choice is unique.

For $\beta$-string, the same argument says that we can put $\cJ$ and
$\tilde\cJ$ on the same sheet of their hyperboloid.  By using $SL(2,\RR)$
we can fix $\cal J$ to, say the point $r=0$ on $S^2_{1}+$.  See
\secr{Hyperboloid} for an explanation of the notation.  The subgroup of
$SL(2,R)$ that leaves $\cal J$ invariant is the compact $U(1)$ generated
by its zero mode.  Its orbits are parameterized by $r$, which measures
the separation between $\cal J$ and $\tilde \cJ$.  It ranges from $0$
to $\infty$.  We call N=2 string theory parameterized by it as type
$\beta_r$.

Starting with the partially reduced parameter space in
\eqr{PartialReducedFlatSpaceChoicesTwoTwo}, what we have just shown
is in algebraic terms
\beq
    \frac {(S^2_{1+} \union S_{1+}^2{}') \times (S^2_{1+} \union S_{1+} ^2{}')}
    {\ZZ_2 \times O(4)}
    = \{\mbox{1 point} \} \union 
    \frac {S^2_{1+} \times {S^2_{1+}}} {SL(2,\RR)}\ .
\eeq 
Again the first part of the union corresponds to the $\alpha$ string; 
the second to the $\beta$ string.
The meaning of the $\ZZ_2$ on the LHS and its effect on 
the RHS exchanging are the same as in the Euclidean case.

\subsection {Toroidal Compactification}
\label {sec:ToroidalCompactification}
Now we go one step up in complexity and consider toroidal compactifications.  The target space remains flat, so the theory remains free.  The chiral algebra therefore does not change, and it still contains the fat, free $N=4$ SCA.  Therefore the analysis of
\secr{NTwoInRFourSimpleRealization} remains applicable,
and  the number of ways of embedding an $N=2$ SCA remains the same.
However while the symmetries acting just on the $\Psi$'s do not change because they are not sensitive to the compactification, rotation symmetries on the $X$'s
is broken by it.  Therefore the number of distinct $N=2$ strings in
general increases.  In particular, for both the left and the right movers, one can still use the rotation symmetries on the fermions to bring $M+M^*$ to $\id$ but the symmetries acting $X$ are reduced because there is less isometry.  This affects the discussion of the space of inequivalent choices of $\cal J$ and $\tilde\cJ$. 
For all cases but the case of total compactification on a sufficiently irregular torus, 
the geometry is invariant under some parity changing rotation, so
the $\ZZ_2$ symmetry that exchanges the two $SU(2)$'s always hold. 
Therefore the theory still has two broad types: $\alpha$ and $\beta$ strings.  We now consider the rest of the story case by case, using the stabilizers of subspaces of
 $\RR^4$ given in \secr{StabilizerEuclidean} and \secr{StabilizerTwoTwo}.
One notes that for a special size and shape of the torus the target
space has nongeneric discrete isometries that reduce the parameter space
of the $N=2$ strings.  For example, if a $n$-torus is rectangular, it will have ${\ZZ_2}^n$ inversion symmetries.  If it is square it will also have power of $\ZZ_4$ symmetries.  
We shall not consider these situations here as
our purpose here is not to exhaustively catalog all compactifications
but rather to give the generic picture and to illustrate the systematic procedures.  The extra symmetries can be
easily taken into account
by the same method and steps given here. A standard but obligatory caveat: when there is a compactified time-like direction it is not clear whether the physics of the theory degenerates.  To be fair, one should first come up with a framework for physics in two times before such issues can even be meaningfully posed for N=2 string theories!

    The rotational symmetry of a generic torus is a total inversion, which we denote here by $\fI$.  So the parameter space of gauging is  
\beq    \label{eq:GeneralParameterSpaceWithNoParityEuclidean}
    \frac {(RP^2 \union RP^2{}') \times (RP^2 \union RP^2{}')}
    {\fI \times \mbox{Isometry of the uncompactified space}}
\eeq
for Euclidean signature, and
\beq    \label{eq:GeneralParameterSpaceWithNoParityTwoTwo}
    \frac {(S^2_{1+}\union S_{1+}^2{}') \times (S^2_{1+}\union S_{1+}^2{}')}
    {\fI \times \mbox{Isometry of the uncompactified space}}
\eeq
for $(2,2)$ signature.  This applies for the generic case when there is 
a physically relevant and \emph {generically} valued background for the 
$B$ field on the torus.  The CFT would then not have any symmetry that 
flips worldsheet parity.  For special cases in which it does, the 
parameter space would instead be 
\beq    \label{eq:GeneralParameterSpaceWithSomeParityEuclidean}
    \frac {(RP^2 \union RP^2{}') \times (RP^2 \union RP^2{}')}
    {\ZZ_2 \times \fI \times \mbox{Isometry of the uncompactified space}}
\eeq
for Euclidean signature, and
\beq    \label{eq:GeneralParameterSpaceWithSomeParityTwoTwo}
    \frac {(S^2_{1+}\union S_{1+}^2{}') \times (S^2_{1+}\union S_{1+}^2{}')}
    {\ZZ_2 \times \fI \times \mbox{Isometry of the uncompactified space}}
\eeq
for $(2,2)$ signature.
The $\ZZ_2$ is the symmetry that flips parity.  
In special cases where the background $B$ vanishes or take special
values so that the worldsheet theory is simply parity invariant,  this
$\ZZ_2$ just exchanges the left and right mover's space of choices.  We
can then evaluate the above one step further and obtain
\beq    \label{eq:GeneralParameterSpaceWithZeroBEuclidean}
    \frac { \left[ \left( RP^2 \times RP^2{}'\right) \union 
    \left( \frac {RP^2 \times RP^2} {\ZZ_2} \union 
    \frac {RP^2{}' \times RP^2{}'} {\ZZ_2}\right) \right]}
    {\fI \times \mbox{Isometry of the uncompactified space}}
\eeq
for Euclidean signature, and
\beq    \label{eq:GeneralParameterSpaceWithZeroBTwoTwo}
    \frac {\left[ \left(S^2_{1+} \times S^2_{1+}{}'\right) \union 
    \left(\frac {S^2_{1+} \times S^2_{1+}} {\ZZ_2} \union 
    \frac {S^2_{1+}{}' \times S^2_{1+}{}'} {\ZZ_2}\right) \right]}
    {\fI \times \mbox{Isometry of the uncompactified space}}
\eeq
for $(2,2)$ signature.

\subsubsection {$\RR^3\times S^1$}
$B$ field background is again irrelevant here.   Therefore the
worldsheet theory has parity symmetry which should be quotiented.

\paragraph {Euclidean Space}
In this case, the rotational isometry is reduced to $O(3)$.  This means
$SU(2)_{O} \times SU(2)_{O'}$ is broken down to a diagonal $SU(2)$ but
the $\ZZ_2$ that exchanges them is not broken.
Therefore one can still bring $\cJ$ to the $\cJ^{[3]}$ for the left
mover.  The $\ZZ_2$ inversion of the $S^1$ does the same thing.  In fact this inversion combined with an inversion of the $R^3$ does nothing to the complex structures.

For type $\beta$, it is the same situation as before, by a
rotation that leave the left mover's $\cJ$ invariant one can reduce
the choice of possibility to $\theta \in [0, \half \pi]$.  For type
$\alpha$, unlike in $\RR^4$, we cannot rotate the $\tilde\cJ$ at will
because only diagonal action of the two $SU(2)$'s are allowed now. 
Therefore the same situation as type $\beta$ obtains: the theory is
parameterized by $\theta \in [0, \half \pi]$.  Indeed, one can see the
two cases more symmetrically.  Because the two $SU(2)$ are now coupled,
one can think of the two complex structures for the left and right
movers as two points on the \emph {same} $RP^2$.  An  arbitrary rotation
can fix one point to be, say, the north pole, then the residual rotation
can fix the longitude of the other point.  The remaining parameter is 
the latitude.  Identifying antipodal points reduce the original $S^2$
to $RP^2$ and the range of latitude to $0 \ldots \half \pi$.  The
same picture applies to type $\alpha$ as well as $\beta$, even though
for the same size of the circle, the two are distinct $N=2$ strings. 
This is not a coincidence.  As will be shown later, they are related by
T-duality on the $S^1$.

In algebraic terms, the parameter space of gauging is calculated from \eqr{GeneralParameterSpaceWithZeroBEuclidean}
\beq
    \frac {\left[ \left(RP^2 \times RP^2{}'\right) \union 
    \left(\frac {RP^2 \times RP^2} {\ZZ_2} \union 
    \frac {RP^2{}' \times RP^2{}'} {\ZZ_2}\right) \right]}
    {\fI \times O(3)} = 
    \frac {RP^2 \times RP^2{}'} {SU(2)} \union
        \frac {RP^2 \times RP^2} {SU(2)}
\eeq
Again the first part of the union corresponds to the $\alpha$ strings; 
the second $\beta$ strings.  Despite the presence of the prime on the 
$\alpha$ part, the two are actually identical since the $SU(2)$ on the 
denominator acts simultaneously on both $RP^2$ and $RP^2{}'$ the same
way.

\paragraph {$(2,2)$ Space}
In this case, the rotational isometry is reduced to $O(2,1)$.  This means
$SL(2,\RR)_{O} \times SL(2,\RR)_{O'}$ is broken down to a diagonal $SL(2,\RR)$ but the $\ZZ_2$ that exchanges them remains unbroken.
Therefore one can still bring $\cJ$ to the $\cJ^{[3]}$ for the left mover. 
For type $\beta$, it is the same situation as before, by a rotation
that leaves the left mover's $\cJ$ invariant one can reduce the choice
of possibility to $\gamma\in [0,\infty]$.  For type $\alpha$, unlike in
$\RR^4$, we cannot rotate the $\tilde\cJ$ at will because only diagonal
actions of the two $SL(2,\RR)$'s are allowed now.  Therefore the same
situation as type $\beta$ obtains: the theory is parameterized by
$\gamma \in [0, \infty]$.  Indeed, one can see the two cases more
symmetrically.  Because the two $SL(2,\RR)$ are now coupled, one can
think of the two complex structures modulo signs  for the left and
right movers as two points on the \emph{same} $S^2_{1+}$.  By an 
$SL(2,\RR)$ action one can fix one point to be at $r=0$.  This leaves
an $U(1)$ rotation that changes $\theta$ but does not affect $r$.  
It can fix the azimuthal angle of the other point.   The
remaining parameter is the $r$ coordinate for the latter, 
which ranges from $0$ to $\infty$. 
The same picture
applies to type $\alpha$ as well as $\beta$, even though for the same
size of the circle, the two are distinct $N=2$ strings.  This is not a
coincidence.  As will be shown later, they are related by T-duality on
the $S^1$.

This discussion essentially parallels that of Euclidean space but with the replacement of $SU(2)$ by $SL(2,\RR)$, $S^2$ by $S^2_1$, and $RP^2$ by $S^2_{1+}$.  So the space of both $\alpha$ and $\beta$ strings is a parameter $r$ ranging from 0 to $\infty$. 
In algebraic terms this means that we have calculated from \eqr{GeneralParameterSpaceWithZeroBTwoTwo} that
\beq
    \frac {\left[ \left(S^2_{1+} \times S^2_{1+}{}'\right) \union 
    \left(\frac {S^2_{1+} \times S^2_{1+}} {\ZZ_2} \union 
    \frac {S^2_{1+}{}' \times S^2_{1+}{}'} {\ZZ_2}\right) \right]}
    {\fI \times O(2,1)} = 
    \frac {S^2_{1+} \times S_{1+}^2{}'} {SL(2,\RR)} \union 
        \frac {S^2_{1+} \times S_{1+}^2} {SL(2,\RR)} \ .
\eeq
Again the first part of the union corresponds to the $\alpha$ strings; 
the second $\beta$ strings.  Despite the presence of the prime on the 
$\alpha$ part, the two are actually identical since the $SL(2,\RR)$ 
on the denominator acts simultaneously on both $S^2_{1+}$ and 
$S^2_{1+}{}'$ the same way.

\subsubsection {$\RR^2 \times T^2$}
Nonvanishing $B$ field background on the Torus begins to play a role
now.  Unless $B = 0$ or $B = \pi$ in some appropriate unit, worldsheet
parity is broken, because it changes the sign of $B$.  For nongeneric
value or $B$ and torus configuration, there is no symmetry in the CFT
that changes the worldsheet parity.  Left and right makes a difference
and we generically should not quotient the parameter space by $\cal J
\leftrightarrow \tilde \cal J$.

\paragraph {Euclidean Space}
In this case the rotational isometry is only $O(2)$, it is embedded in
the $O(3)$ above as the stabilizer of an 3-vector.  
The parity changing elements in $O(2)$ again ensure 
the dichotomy into $\alpha$ and $\beta$ types as before.  
Also as before, we can 
visualize this as two points on the same $RP^2$.  
Breaking $O(3)$ down to $O(2)$ amounts to choosing a point on the
$S^2$.  The inversion $\fI$ inverts that point to its antipode and thus
quotienting by it turns the point on $S^2_1$ into a point on $S^2_{1+}$.
 Let it be the north pole.  The residual $U(1)$ allows us to fix the
longitude of one of the two points.  The system is therefore naturally
parameterized by two latitudes and one longitudes.  The range of
longitude is $0\ldots 2\pi$, and that of latitude $0\ldots \pi$. 
This is the final answer if the two points are distinct.  Since they 
correspond to $\cal J$ and $\tilde \cal J$ respectively, whether they 
are distinct or not depends on whether we have worldsheet parity
symmetry.   Generically we don't, as it would change the sign of the $B$
field.   If we do have parity symmetry because $B=0$ or $B=\pi$ in
some  appropriate unit, we have to further quotient the parameter space
by the $\ZZ_2$ that exchanges $\cal J$ and $\tilde\cal J$.  This means
that we can assume a fixed ordering of for example the two latitudes, 
say $\theta_1 \geq \theta_2$.  

Algebraically, we have just evaluated  \eqr{GeneralParameterSpaceWithNoParityEuclidean}
\beq    \label {eq:ParameterSpaceTTwoEuclideanGeneric}
    \frac {(RP^2 \union RP^2{}') \times (RP^2 \union RP^2{}')}
    {\fI \times O(2)} 
    = \frac {RP^2 \times RP^2{}'} {U(1)} \union
        \frac {RP^2 \times RP^2} {U(1)}
\eeq
for the generic case and \eqr{GeneralParameterSpaceWithZeroBEuclidean}
\beq    \label {eq:ParameterSpaceTTwoEuclideanParity}
    \frac {\left[ \left(RP^2 \times RP^2{}'\right) \union 
    \left(\frac {RP^2 \times RP^2} {\ZZ_2} \union 
    \frac {RP^2{}' \times RP^2{}'} {\ZZ_2}\right) \right]}
    {\ZZ_2 \times \fI \times O(2)} = 
    \frac {RP^2 \times RP^2{}'} {\ZZ_2 \times U(1)} \union
        \frac {RP^2 \times RP^2} {\ZZ_2 \times U(1)}.
\eeq
for the parity invariant case.
Again the first part of the union corresponds to the $\alpha$ strings; 
the second $\beta$ strings.  The story about the prime is also 
unchanged: despite it, $\alpha$ and $\beta$ parts are identical.

\label{page:TDualPuzzleOne}
As we have already mentioned several times and we will show in
\secr{TDuality}, $\alpha$ and $\beta$ strings are related by
T-duality.  We have just see this reflected in identical parameter
space of gauging for $\alpha$ and $\beta$ in $S^1$ compactification. 
The situation in $T^2$ is analogous but more subtle.  The two parts are
identical both for the generic case and for the parity invariant case. 
At first sight there is actually puzzling, because if we start with the
nongeneric case of vanishing $B$ on the torus, and suppose the torus is
non rectangular, then T-dualizing along one of the periodically
compactified direction will lead to another torus with nonvanishing $B$
field.  If this is na\"\i{}vely thought as corresponding to the generic
case one would have obtained a contradiction with T-duality as the
parameter space of gauging would have differed by a factor of two
between the two sides.  However this thinking is incorrect.  T-duality
is an equivalence relation between different target space geometries
and does not change the underlying CFT.  In particular it cannot remove
or create a symmetry.  If the original theory is parity symmetric, its
T-dual would also have it.  However, since T-duality rotate the labels
of the right moving fields, the worldsheet parity changing symmetry of
the T-dual theory would have to involve a similar rotation between the
labels of the left and right moving fields in addition to swapping
them.  With this taken into account, the parameter space of gauging of
the T-dual theory is again \eqr{ParameterSpaceTTwoEuclideanParity}
instead of \eqr{ParameterSpaceTTwoEuclideanGeneric} and there is no
contradiction.

\paragraph {$(2,2)$ Space with spatial $T^2$}
In this case the rotational isometry is only $O(2)$.  It is embedded in
the $O(2,1)$ above as the stabilizer of a space-like 3-vector 
(i.e. associated with the (1) of (2,1)).  The parity changing elements
in $O(2)$ again ensure the dichotomy into $\alpha$ and $\beta$ types as
before.  Also as before, we can visualize this as two points on the same
$RP^2$. Breaking $O(2,1)$ down to $O(2)$ amounts to choosing a point
on the $S^2_1$.  The inversion $\fI$ inverts that point to the other
sheet and thus quotienting by it turns the point on $S^2_1$ into a point
on $S^2_{1+}$.  Let it be the point $r = 0$.  The residual symmetry
allows us to fix the azimuthal angle of one of the two points.  The
system is therefore naturally parameterized by one azimuthal angle and
two radial parameters. The range of the former is $0\ldots 2\pi$, and
that of later $0\ldots \infty$.  
This is the final answer if the two points are distinct.  Since they 
correspond to $\cal J$ and $\tilde \cal J$ respectively, whether they 
are distinct or not depends on whether we have worldsheet parity
symmetry.   Generically we don't, as it would change the sign of the
$B$ field.   If we do have parity symmetry for a particular
configuration,  we have to further quotient the parameter space by the
$\ZZ_2$ that exchanges $\cal J$ and $\tilde \cal J$.  This means that
we can assume a fixed ordering of for example the two radial 
parameters, say $r_1 \geq r_2$.  

Algebraically, we have just evaluated  \eqr{GeneralParameterSpaceWithNoParityTwoTwo}
\beq    \label {ParameterSpaceTTwoTwoTwoGeneric}
    \frac {(S^2_{1+} \union S^2_{1+}{}') \times (S^2_{1+} \union S^2_{1+}{}')}
    {\fI \times O(2)} 
    = \frac {S^2_{1+} \times S^2_{1+}{}'} {U(1)} \union
        \frac {S^2_{1+} \times S^2_{1+}} {U(1)}
\eeq
for the generic case and \eqr{GeneralParameterSpaceWithZeroBTwoTwo}
\beq    \label {eq:ParameterSpaceTTwoTwoTwoParity}
    \frac {\left[ \left(S^2_{1+} \times S^2_{1+}{}'\right) \union 
    \left(\frac {S^2_{1+} \times S^2_{1+}} {\ZZ_2} \union 
    \frac {S^2_{1+}{}' \times S^2_{1+}{}'} {\ZZ_2}\right) \right]}
    {\ZZ_2 \times \fI \times O(2)} = 
    \frac {S^2_{1+} \times S^2_{1+}{}'} {\ZZ_2 \times U(1)} \union
        \frac {S^2_{1+} \times S^2_{1+}} {\ZZ_2 \times U(1)}.
\eeq
for the parity invariant case.
Again the first part of the union corresponds to the $\alpha$ strings; 
the second $\beta$ strings.  The story about the prime is also 
unchanged: despite it, $\alpha$ and $\beta$ parts are identical.  
As the discussion about T-duality is the same we shall not repeat it.

\paragraph{$(2,2)$ Space with one space and one time compactified}
In this case the rotational isometry is only $O(1,1)$. It is embedded in
the $O(2,1)$ above as the stabilizer of an 3-vector.  The parity
changing elements in $O(2)$ again ensure the dichotomy into $\alpha$ and
$\beta$ types as before.  We can again visualize this as two points on
the same $S^2_{1+}$. Breaking $O(2,1)$ down to $O(1,1)$ amounts to
choosing direction in the xy-plane and stratify $S^2_1$ into plane
sections perpendicular to that direction and hence $S^2_{1+}$.  Let us
choose the direction be along $y$. The inversion $\fI$ is not relevant
this time. The residual Abelian symmetry acts transitively on the plane
sections of $S^2_{1+}$ as a Lorentz boost, with say $z$ being the
``time.''  This allows us to fix the $x$-coordinate of one of the two
points to for example $0$.  The system is therefore naturally
parameterized by one $x$ and two $y$ coordinates. There range would be
all from $0\ldots \infty$.
This is the final answer if the two points are distinct.  Since they 
correspond to $\cal J$ and $\tilde\cal J$ respectively, whether they 
are distinct or not depends on whether we have worldsheet parity
symmetry.   Generically we don't, as it would change the sign of the
$B$ field.   If we do have parity symmetry for a particular
configuration,  we have to further quotient the parameter space by the
$\ZZ_2$ that exchanges $\cal J$ and $\tilde\cal J$.  This means that
we can assume that the unfixed $x$-coordinate is positive.  

Algebraically, we have just evaluated  \eqr{GeneralParameterSpaceWithNoParityTwoTwo}
\beq    \label {eq:ParameterSpaceTTwoTwoTwoMixGeneric}
    \frac {(S^2_{1+} \union S^2_{1+}{}') \times (S^2_{1+} \union S^2_{1+}{}')}
    {\fI \times O(1,1)} 
    = \frac {S^2_{1+} \times S^2_{1+}{}'} {\RR} \union
        \frac {S^2_{1+} \times S^2_{1+}} {\RR}
\eeq
for the generic case and \eqr{GeneralParameterSpaceWithZeroBTwoTwo}
\beq    \label {eq:ParameterSpaceTTwoTwoTwoTwoMixParity}
    \frac {\left[ \left(S^2_{1+} \times S^2_{1+}{}'\right) \union 
    \left(\frac {S^2_{1+} \times S^2_{1+}} {\ZZ_2} \union 
    \frac {S^2_{1+}{}' \times S^2_{1+}{}'} {\ZZ_2}\right) \right]}
    {\ZZ_2 \times \fI \times O(1,1)} = 
    \frac {S^2_{1+} \times S^2_{1+}{}'} {\ZZ_2 \times \RR} \union
        \frac {S^2_{1+} \times S^2_{1+}} {\ZZ_2 \times \RR}.
\eeq
for the parity invariant case.
Again the first part of the union corresponds to the $\alpha$ strings; 
the second $\beta$ strings.  The story about the prime is also 
unchanged: despite it, $\alpha$ and $\beta$ parts are identical.  
As the discussion about T-duality is the same we shall not repeat it.

\subsubsection {$\RR^1 \times T^3$}

\paragraph {Euclidean Space}
In this case the only ``rotation'' isometry of the noncompact space is
is its $\ZZ_2$ inversion.  The joint action of that inversion and $\fI$
has no effect on the complex structures.  By themselves either can be
used to show that the same $\alpha$-$\beta$ dichotomy obtains.  The
parameter space is that of two points on $RP^2$ with no
symmetry to further reduce.  
Generically this is the final answer.  For special value of the $B$
field on the torus and suitable torus configuration, the theory can
have a worldsheet parity changing symmetry and the parameter space of
gauging becomes two identical points on $RP^2$.
Algebraically, this gives
\beq
        \left(RP^2 \times RP^2{}' \right) \union 
                \left(RP^2 \times RP^2\right)
\eeq
for the generic case and 
\beq
\frac {RP^2 \times RP^2{}'} {\ZZ_2} \union
        \frac {RP^2 \times RP^2} {\ZZ_2}
\eeq
for the parity symmetric case.

\paragraph {$(2,2)$ Space}
The same statement about the isometries for the Euclidean space also
applies here.  The parameter space is that of two points on a
$S^2_{1+}$.  Generically the two points are distinct, but for cases 
where the CFT is parity invariant, they are identical.

\subsubsection {$T^4$}
In this rather special case, there is no compact directions left.
Possible physical pathology aside, we can discuss the parameter space
of gauging as it is.  If the torus is the direct product of
metric spaces in the form $S^1 \times T^3$ the answer is the same
as the one for $\RR \times T^3$ derived above because they have 
the same set of isometries.

If instead the $T^4$ is a generic one, the only isometry left is $\fI$,
the total inversion of $T^4$.  It has no effect on the complex
structure.  
The algebraic result for the generic case is just 
\eqr{PartialReducedFlatSpaceChoicesEuclidean}:
\beq    \label {eq:TFourEuclideanGeneric}
    \left(RP^2 \times RP^2{}' \union RP^2{}' \times RP^2\right) \union 
     \left(RP^2 \times RP^2 \union RP^2{}' \times RP^2{}'\right).
\eeq
for Euclidean signature and 
and     \label {eq:TFourTwoTwoGeneric}
\eqr{PartialReducedFlatSpaceChoicesTwoTwo}:
\beq
    \left(S^2_{1+} \times S^2_{1+}{}' \union S^2_{1+}{}' \times S^2_{1+}\right) \union 
     \left(S^2_{1+} \times S^2_{1+} \union S^2_{1+}{}' \times S^2_{1+}{}'\right).
\eeq
for $(2,2)$ signature.
When the $B=0$ or other special values such that the CFT 
is parity invariant, we should further reduce by $\ZZ_2$ and obtains 
\beq    \label {eq:TFourEuclideanParity}
   (RP^2 \times RP^2{}') \union \left( \frac {RP^2 \times RP^2} {\ZZ_2} 
        \union \frac {RP^2{}' \times RP^2{}'} {\ZZ_2} \right)      
\eeq
for Euclidean signature and 
\beq    \label {eq:TFourTwoTwoParity}
    (S^2_{1+}\times S_{1+}^2{}') \union \left( 
        \frac {S^2_{1+}\times S_{1+}^2} {\ZZ_2} \union
        \frac {S^2_{1+}{}'\times S_{1+}^2{}'} {\ZZ_2} \right) 
\eeq
for $(2,2)$ signature.  
$\alpha$ string correspond to the first pair of parentheses 
and $\beta$ the second.  
Note that there are disjoint subclasses with one or both types now. 
This is because we no longer have a target space parity changing
symmetry to bring $\cal J$ to a particular $S^2$ or $S^2_1$.  
Put it in a geometric way,  a generic $T^4$ does not treat the two
classes of complex structures symmetrically.   

Now
\eqr{TFourEuclideanParity} and \eqr{TFourTwoTwoParity} raise a puzzle
similar to the one discussed on \pageref{page:TDualPuzzleOne} and it has a
similar resolution.  The parameter space of gauging for $\alpha$ and
$\beta$ strings have the same ``size,'' but are very different
topologically spaces.  In particular $\alpha$ string just has one
component while $\beta$ has two disjoint ones.  There is no reason 
why the latter two should be joined as there is no 
continuous transition from one to the other.  Does this contradict
T-duality?  Of course not.  As we discussed in the last puzzle, the
T-dual theory has a worldsheet parity changing symmetry that also
involves relabeling the fields.  That relabeling is itself like a
parity operation on the target space.  This is what happens with
T-duality along a single direction and this is also precisely why it
relates $\alpha$ and $\beta$ strings: it swap the two $S^2$'s or
$S^2_1$'s.  Now quotient \eqr{TFourEuclideanGeneric}  and
\eqr{TFourTwoTwoGeneric} by this $\ZZ_2$ and we get again
\eqr{TFourEuclideanParity} and \eqr{TFourTwoTwoParity} respectively,
with one crucial difference: the parameter spaces for 
$\alpha$ and $\beta$ are
swapped.  Therefore there is no contradiction with T-duality.

Note that a generic $T^4$ means all
directions, include both ``time'' directions, are mixed with each other
by the metric and periodicity conditions.  While geometrically this is
not unusual, 
its physical relevance and implication is far from obvious.  
But the reader has already been amply warned.

\subsection {General Dichotomy}
\label {sec:alphabetaSynthesis}
We have shown that in all possible 4d flat space, $N=2$ strings come in
two broad flavors, $\alpha$ and $\beta$ types, each in turn usually
contains a continuous family.  This seems very reminiscent of the
dichotomy between type IIA and IIB superstrings.  Indeed, as we shall
see, when compactified on a circle, $\alpha$ and $\beta$ $N=2$ strings
are T-dual to each other, much as type IIA and IIB superstrings.  A
question then naturally arises as to whether this dichotomy is special to
flat space or can be generalized to other spacetime geometry. 
Here we give an affirmative answer.

Supersymmetric $\sigma$-models can be conveniently written in terms of 
$N=1$ superspace\footnote 
    {For a brief review of $N=1$ superspace in 2d, see, for example, \S
    4.1.2 of \cite{Green:sp}.}:
\beq    \label {eq:NIsOneSigmaAction}
        S = \myI {d^2\sigma d^2\theta} 
        (\cal G_{IJ} + \cal B_{IJ}) \bD \bX^I \tilde D \bX^J \ .
\eeq

As written, it is already supersymmetric: there is no condition on the
geometric quantity $\cal G$ and $\cal B$.  For this reason, it is
believed that all $\sigma$-models with $N=1$ supersymmetry can be
described by \eqr{NIsOneSigmaAction}, including those with extended
supersymmetry, as extended supersymmetry algebra always include the
$N=1$ supersymmetry algebra as a proper subalgebra.  
Therefore the following discussion and results, in particular the 
appearance of two almost complex structures, are not made less general 
by the use of $N=1$ superspace, which is just for convenience. 
In \cite{Gates:1984nk},
$N=2$ (as well as $N=4$) $\sigma$-models were considered in this framework by introducing
a second pair of supersymmetry transformations defined by\footnote{
        We note that the authors of \cite{Gates:1984nk} considered
compact $N=2$ supersymmetry, in sense that the R-symmetry group is a
compact $U(1)$.  That is precisely the case relevant to this paper.}
\beq
        \bQ_2 \bX^I = {\cJ^I}_J \bD \bX^J\ , \csp 
        \tilde \bQ_2 \bX^I = {\tilde \cJ^I}_{\;\;\; J} \tilde \bD \bX^J \ .
\eeq
They derived from $N=2$ supersymmetry algebra that both $\cJ$ and
$ \tilde\cJ$ are integrable complex structures.  They also found
that the invariance of the action \eqr{NIsOneSigmaAction} under the
second supersymmetry implies that both complex structures make $\cal G$ hermitian and that they are respectively covariantly constant:
\beq
        D \cJ  = \tilde D  \tilde\cJ = 0 \ .
\eeq
The caveat is that $D$ and $\tilde D$ are two covariant derivatives 
defined respectively with two affine connections given by 
\beq
        \Gamma = \Gamma_0 + \cal H\ , \csp 
       \tilde \Gamma = \Gamma_0 - \cal H\ ,
\eeq
where all (implicit) indices are downstairs and $\Gamma_0$ 
is the Levi-Civita connection associated with $\cal G$.  
  From this they constrains 
$\cJ$ and $\tilde \cJ$ and determine $\cal H$ in terms of them.

Now for a given $\sigma$-model, there can be no, one or more than one
candidate for $\cal J$ and $ \tilde\cal J$, depending on the number of
extended supersymmetry the theory has. As a pre-theory for $N=2$ string,
we require there to be at least one candidate for $\cal J$ and one for
$ \tilde\cal J$.  Though the simplest and most widely studied cases have
$\cal J =\tilde \cal J$, it does not have to be so \cite{ Gates:1984nk}. There
can also be more than one choices for $\cal J$ and $ \tilde\cal J$.  If
so the extended supersymmetry is more than $N=2$. The choice is
irrelevant for the $\sigma$-model itself, other than that their
existence implies the existence of a bigger symmetry group and $N=2$
supersymmetry and $U(1)$ R-symmetry.  However, for the $N=2$ string the
$\sigma$-model is only the pre-theory.  To specify the string theory, we
have to make a specific choice for $\cal J$ and $ \tilde\cal J$
respectively.  This choice determines the action of the $U(1)$
R-symmetry and hence the left and right affine $U(1)$ currents of the
$N=2$ SCA to be gauged.  Different choices lead to different $N=2$
string theory.  In \secr
{AlphaBeta} we have seen this to happen when the target space is flat, 
and we have
examined the space of physically inequivalent choices.  We shall return to examine cases of curved target spaces
in \secr {CurvedSpace}.  

This choice allows us to generalize the $\alpha$ and $\beta$ dichotomy
to arbitrary spacetime $\fX$ described by $N=2$ $\sigma$-models.  As
shown in \secr{HermitianStructureEuclidean} and
\secr{HermitianStructureTwoTwo}, the space of complex structures at
each point in a $4d$ space is the disjoint union of two identical
manifolds.  For Euclidean metric, it is two $S^2$; for $(2,2)$ metric,
it is two $S^2_1$.  A complex structure for the whole manifold is a
section of a fibration of this disjoint union over $\fX$.  This
fibration is completely determined by the topology of the target space 
manifold, as the transition function for the fibers can be related to
that of the tangent bundle, which is either $O(4)$ or $O(2,2)$
depending on the signature of the metric.  Special orthogonal elements
acts on the two component manifolds separately; elements that change
the orientation of $\RR^4$ map one to the other $S^2$ or $S^2_1$.  
While locally on each coordinate patch one can distinguish the two
manifolds, whether this distinction can be extended globally over the
whole target space is a different matter.  

Abstractly speaking this labeling yields a vector bundle over the target space whose fiber is the vector space $\ZZ_2$ understood as a module of the Abelian group of $\ZZ_2$.  Its transition function between two patches is the sign of the determinant of the corresponding transition function of the tangent bundle.  This $\ZZ_2$ bundle is trivial if and only if the first Stiefel-Whitney class of $\fX$ vanishes, i.e. $\fX$ is orientable.  
On the other hand, nonorientability is an obstruction to the existence of an almost
complex structure because parity changing maps are not holomorphic.  
For example, on a 2d surface reflection along the
$dz + d\bar z$ direction exchanges $dz$ and $d\bar z$.  Any other
reflection (parity) along a certain direction does the same with some
additional phase factor.  Therefore one cannot globally distinguish $dz$
and $d\bar z$, which is what an almost complex structure does, if one
cannot eliminate all reflections from the transition functions of tangent
bundles across patches, which is what nonorientability means.  This
argument easily generalizes to higher dimensions.  Therefore an almost
complex manifold is necessarily orientable, and so we should restrict
ourselves to this case since $\fX$ is almost complex with 
$\cal J$ and $ \tilde\cal J$.

On orientable $\fX$, the two locally disjoint components of the fiber
are also globally distinct.  The fibration therefore also splits into
two components.  Each one is a fibration of $S^2$ or $S^2_1$ over the
target space.  In defining the $\sigma$-model\eqr{NIsOneSigmaAction}, we
have to make a choice for $\cJ$ and another for $ \tilde\cJ$.  There are
in general many choices.  The first level of classification is given by
the discrete choices of which components of the fibration they
respectively belong. 
\begin{quote}
\emph{We define $N=2$ $\alpha$ string as those 
defined with $\cJ$ and $ \tilde\cJ$ from different
components.  $N=2$ $\beta$ strings, on the other hand, 
have them from the
same component.}
\end{quote}

\section {$N=2$ Strings in Curved Spaces}
\label {sec:CurvedSpace}

Flat space is not the only space in which $N=2$ string can propagate. 
Since any $N=2$ SCFT with the appropriate central charge is a starting
point for constructing a $N=2$ string theory, it is interesting and
imperative to investigate the case of interacting $N=2$ SCFTs.  Some of
them have spacetime interpretation, being $N=2$ superconformal invariant
$\sigma$-models.  These have been studied intensively, mainly because
$N=2$ SCFT provides supersymmetric compactification for the superstring
theories.  They can be used for $N=2$ string if their central charge is
$6$, which usually translates to $4$ real dimensions.  To our knowledge,
reasonably rigorous analysis that completely classifies or characterizes such
spaces without extra (hidden) assumption does not yet exist.  Instead, there are
fairly large class of examples characterized by some additional
geometric conditions such as holonomy, torsion, and superspace
representations.  In this section, we shall consider several of them. 
The aim is to solve for some families of $N=2$ string theories one can
define in these spaces.  The principles of such analysis have already
been laid down in \secr {Recipe}.  We
restrict ourselves to simple embeddings of $N=2$ SCAs in the $N=4$ SCA, 
to defined shortly,
that is characteristic of all the examples we shall study.  We first
establish a general result about all such embeddings.  Then we remove
the degeneracy among them caused by the symmetry of the corresponding
field theory.  There are interesting examples of curved spaces in which $\alpha$ string propagate, but here we concentrate on $\beta$
strings.  Unlike the case of flat space, they generally
do not coexist in the same curved spacetime.  This does not, however,
mean that they are unrelated, but we shall wait until 
\secr{TDuality} to see the
relation between them.  Both $(4,0)$ and $(2,2)$ signatures are
considered.

The recipe in \secr {Recipe} requires us to find out all such
symmetries given that we have all the information about the geometry of
$\fX$.  If we also have enough control over quantum corrections, we can
also find out how at the quantum level they act on the realizations of
the $N=2$ SCA's we want to gauge.  Sometimes, however, we are just
given with some general features of the pre-theory, such as the presence 
of a bigger symmetry algebra $\mathfrak A$.  
For the examples in this section $\mathfrak A$
is the $N=4$ SCA.  We then have to make
the assumption of genericness that this is all the symmetry the theory
has, which is already is quite informative.  In lieu of the assumption
of simple realization in terms of fundamental matter fields
\eqr{SimpleNTwoRealization}, we solve for all ``simple'' embeddings of the 
$N=2$ SCA in $N=4$ SCA.  They are 
related to the automorphisms of group of $\mathfrak A$, the detail
depends on $\mathfrak A$.  That has to be quotiented by the symmetry of
the theory, which at least include the inner automorphisms of $N=4$
SCA.  For other chiral algebra $\mathfrak A$, 
the procedure would be analogous.

\subsection {Simple Embeddings of $N=2$ SCA in $N=4$ SCA}

\subsubsection {Compact $N=4$}

Define a simple embeddings of a compact $N=2$ SCA in a compact $N=4$ SCA by 
\beq    \label {eq:SimpleEmbeddingNTwoInNFourCompact}
        G^+ = m_{\alpha\beta} G^{\alpha,\beta}, \csp 
        G^- = - m^*_{\alpha\beta} {(\sigma^2)^\alpha}\gamma 
                {(\sigma^2)^\beta}_\lambda G^{\gamma,\lambda} \ .
\eeq
We can write $m$ in terms of Pauli's $\sigma$ matrices 
(see the appendix for our conventions):
\beq
        m_{\alpha\beta} = 
        (m_0 \epsilon  + \imath m_i \sigma^i)_{\alpha\beta} \ .
\eeq
Note that the $SU(2)_O\times SU(2)_I$ symmetry acts on $m$ by
rotating $m^I$ as a 4-vector under $SO(4)$.  We want $G^\pm$ has the
same OPE as \eqr {NTwoOPE}, with $J$ given by 
\beq
        J = 2 a_i J^i \ .
\eeq
This problem is similar to the one we have solved in \secr
{NTwoInRFourSimpleRealization}.  By a similar analysis, we find that
\beq
        m_I = \half (p_I + \imath q_I), \csp p_I, q_I \in \RR
\eeq
such that 
\beq    \label {eq:pqNormCompact}
       <p,p> = <q,q> = 1\ , \csp <p,q> = 0 \ ,
\eeq
and 
\beq
        -q_0 p_k + p_0 q_k + \epsilon^{ijk} p_i q_j = a_k.
\eeq
We could continue to solve these and other equations resulting from 
requiring \eqr{NTwoOPE}.  Instead, let us note that given \eqr{pqNormCompact},
we can, by an $SO(4)$ rotation, bring $p$ and $q$ to a preferred
form:
\beqar
        p_I = 0, I=1,2,3; &\csp& p_2 = 1;\nono
        q_I = 0, I=0,1,3; &\csp& q_1 = - 1 \ .
\eeqar
Then, 
\beq
        a_3 = 1, \csp a_1 = a_2 = 0 \ ,
\eeq
and
\beq
        G^+ = G^{1,1},\csp G^- = G^{2,2},\csp J = 2 J^3.
\eeq
Thus we have proved that every simple
embeddings of $N=2$ SCA is related to
\eqr{ExampleOfNTwoInNFourCompact} by some action of $SU(2)_O\times SU(2)_I$ automorphism.

Hence in order to find the parameter space of gauging we should start
with $SU(2)_O\times SU(2)_I$.  Right away we can drop $SU(2)_I$
because, as internal automorphism, they are necessarily symmetry of the
CFT.   Furthermore, 
a $U(1)$ subgroup of $SU(2)_O$ just changes
the phase $G^{\pm}$ in opposite directions.  They do not give different
embeddings and should also be quotiented out.
Therefore we get $SU(2)/U(1) \simeq S^2$.  
Finally, there is a $\ZZ_2$ operation in  $SU(2)_I\times SU(2)_O$
that reverses the sign of $J$ and interchanges
$G^{1,1}$ and $G^{2,2}$.  This $\ZZ_2$ is the Weyl group in each $SU(2)$ and 
it also leads to the same $N=2$ SCA.  
Indeed it corresponds to the conjugation automorphism of the latter. 
We are now left with
\beq
\frac{S^2}{Z_2} = RP^2\ .
\eeq
At least part of the symmetries of 
the theory have already been taken into account to arrive at this result.

\subsubsection {Noncompact $N=4$}

Define a simple embeddings of a compact $N=2$ SCA in a noncompact $N=4$ SCA by 
\beq    \label {eq:SimpleEmbeddingNTwoInNFourNoncompact}
        G^+ = m_{\alpha\beta} G^{\alpha,\beta} \ , \csp 
        G^+ = m^*_{\alpha\beta} G^{\alpha,\beta} \ .
\eeq
We can write $m$ in terms of sigma matrices (see the appendix for our notations):
\beq
        m_{\alpha\beta} = 
        (m_0 \epsilon + \imath m_2 \sigma^2 
        + m_1 \sigma^1 + m_3 \sigma^3)_{\alpha\beta} \ .
\eeq
Note that the $SL(2,\RR)_O\times SL(2,\RR)_I$ symmetry acts on $m$ by
rotating $m^I$ as a 4-vector under $SO_0(2,2)$.  We want $G^\pm$ has the
same OPE as \eqr {NTwoOPE}, with $J$ given by 
\beq
        J = 2 a_i J^i \ .
\eeq
This problem is similar to the one we have solved in \secr
{NTwoInRFourSimpleRealization}.  By a similar analysis, we find that
\beq
        m_I = \half (p_I + \imath q_I), \csp p_I, q_I \in \Re
\eeq
such that 
\beq    \label {eq:pqNormNoncompact}
       <p,p> = <q,q> = 1\ , \csp <p,q> = 0 \ ,
\eeq
and 
\beq
        q_0 p_k - p_0 q_k + \varepsilon^{ij}_k p_i q_j = a_k \ .
\eeq
We could continue to solve these and other equations resulting from 
requiring \eqr{NTwoOPE}.  Instead, let us note that given \eqr{pqNormNoncompact},
we can, by an $SO_0(2,2)$ rotation, bring $p$ and $q$ to a preferred
form:
\beqar
        p_I = 0, I=1,2,3; &\csp p_0& = 1;\nono
        q_I = 0, I=0,1,3; &\csp q_2& = \mp 1.
\eeqar
Then, 
\beq
        a_2 = \pm 1, \csp a_1 = a_3 = 0 \ .
\eeq
The choice of the sign cannot yet be fixed  because
flipping it while keep $p_0 = 1$ requires the use of an improper element
of $SO(2,2)$.  However it does not matter, because flipping that sign
correspond to reversing $J$ and interchanging $G^{\pm}$.  That does not
give us a different $N=2$ SCA. So we make a choice and find that  
$G^\pm$ and $J$ are given as in \eqr{ExampleOfNTwoInNFourNoncompact} 
and \eqr{ExampleOfNTwoInNFourNoncompactJ} respectively.
Thus we have proved that every simple
embeddings of $N=2$ SCA is related to
\eqr{ExampleOfNTwoInNFourNoncompact} by some action of $SL(2,\RR)_O\times SL(2,\RR)_I$
automorphism.

Hence in order to solve for the parameter space of gauging we should
start with $SL(2,\RR)_O\times SL(2,\RR)_I$  Right away we can drop $SU(2)_I$
because, as internal automorphism, they are necessarily symmetry of the
CFT.   Furthermore, 
a $U(1)$ subgroup of $SL(2,\RR)_O$ just changes
the phase $G^\pm$ in opposite directions.  It leads to the same 
$N=2$ SCA and thus should be quotiented out.  
We are now left with
\beq
        SL(2,\RR)/U(1) \simeq S^2_1\ .
\eeq
Finally, there is a $\ZZ_2$ operation on $SL(2,\RR)_I\times SL(2,\RR)_O$
that reverses the sign of $J$ and interchanges
$G^{1,1}$ and $G^{2,2}$.  This $\ZZ_2$ is the Weyl group in each 
$SL(2,\RR)$ and the same as the improper element of $SO(2,2)$ 
mentioned above.  
It also leads to the same $N=2$ SCA.  
Indeed it corresponds to the conjugation automorphism. 
We are now left with
\beq
\frac {S^2_1}{Z_2} = S^2_{1+}\ ,
\eeq
At least part of the symmetries of 
the theory have already been taken into account to arrive at this result.

\subsubsection {$N=2$ in Fat, Free $N=4$, a second look}

Before going into the curved spaces we shall do a warm-up exercise by
applying the results above to the case of flat space and recover the same
results as obtained there from free fields.  In fact, this subsection 
and \secr{NTwoInRFourSimpleRealization} can be read independently.

The various automorphism and symmetry of the fat, free $N=4$ SCA
discussed in \secr{FatFreeNIsFour} are clearly related to the symmetry of 
$\RR^4$, discussed in detail in \secr {GeometryOfRFour}. 
Now consider a $\sigma$-model defined on this $\RR^4$,  
the chiral algebra of which has been discussed in 
\secr {FatFreeNIsFour}. 
The rotational
symmetry can act on the bosons $X^I$'s and fermions $\Psi^I$'s.  Because
the bosons and fermions decouple from each other, one can in fact
rotate them independently.  Furthermore, because the left and right
moving chiral fermions are actually different fields, one can also
rotate them further.  This means that in fact there are left and right
chiral affine $SU(2)$ or $SL(2,\RR)$ currents corresponding to them.  
They are the
$J^i$, $J'^i$, and their right chiral counterparts $\tilde J^i$ and
$\tilde J'^i$, in \eqr{PairOfSUTwoCurrent}
 or \eqr{PairOfSLTwoCurrent}.  The bosons, however, cannot
be completely separated into independent left and right moving parts.
The space being noncompact means that the left and right movers share
the same zero modes.  Therefore rotation of the bosons leads to just one
rotation symmetry group that rotates the $X$, which does not correspond
to any chiral current.  The same story applies to parity operations 
that swaps the two spinor indices.  
For fermions each leads to two symmetries, one
for each chirality.  For bosons there is just one acting on
$X$.

Even though there is no symmetry rotating the left and right moving parts
of $X$ independently, the chiral algebra \eqr {RFourFreeFieldOPE} \emph
{is} invariant under independent rotation of $\pa X$ and $\bar \pa X$. 
This is an example where an automorphism of the chiral algebra does not
correspond to a symmetry of the theory.  Therefore it is necessarily an
outer automorphism.  
The origin of all
the symmetry and automorphism of the fat, free $N=4$ SCA should be
evident now.

Given the structure of the fat, free $N=4$ SCA, we learned from 
\secr{FatFreeNIsFour} that there are at
least 4 distinct ways to embed an $N=4$ SCA, depending on which of the
fours quadruplets of supercharges, $G_{00}$, $G_{01}$, $G_{10}$, and
$G_{11}$, one choose.  By a parity operation on $\Psi$, these choices
can be narrowed down between $G_{00}$ and $G_{10}$.
After having chosen which way to embed the $N=4$ SCA, there remain a
continuum of choices to finally fix the $N=2$ SCA embedding.
As discussed \secr {NIsFour}, if the affine group is $SU(2)$, 
the space of different simple embeddings mod $SU(2)_I$ is $RP^2$; 
if the affine group is $SL(2,\RR)$ instead, the space of embeddings 
mod $SL(2,\RR)$ is 
$S^2_{1+}$.  We have seen in \secr {FatFreeNIsFour} 
that these two possibilities correspond to $(4,0)$ and $(2,2)$ metrics 
respectively.
Now putting the left and right movers together and 
we have 
\beq
    ((RP^2 \union RP^2{}') \times (RP^2 \union RP^2{}'))
\eeq
and
\beq
    ((S^2_{1+}\union S_{1+}^2{}') \times (S^2_{1+}\union S_{1+}^2{}')) 
\eeq
for the two types of signatures respectively.  They are identical to
\eqr{PartialReducedFlatSpaceChoicesEuclidean} and 
\eqr{PartialReducedFlatSpaceChoicesTwoTwo}. One can then continue with
the analysis of flat spaces in \secr{RFour} and
\secr{ToroidalCompactification}.  
We remark that when the left and right chiral $N=2$
supercurrents come from
$G_{00}$ and $\tilde G_{10}$ respectively, 
we have an $\alpha$-string.  
When the left and right chiral $N=2$
supercurrents both come from
$G_{00}$, we have a $\beta$-string.

\subsection {$\beta$-string in Curved Space}

As defined in \secr{alphabetaSynthesis}, 
$N=2~\beta$ string is describable by a $\sigma$ model with $N=2$ such that the left and right movers use the same complex structure.

\paragraph {K3}
\label  {sec:KThree}
A generic $K3$ has no isometry.  
Because the
worldsheet field theory is generically not free, we cannot apply the
analysis of 3.2.  That would of course be possible for orbifold limits 
of $K3$, but here we restrict ourselves to the generic case.
Fortunately, it is believed that the supersymmetric $\sigma$-model with 
$K3$ as target space is in fact a $N=4$ SCFT \cite{KThreeExact}.  
For the generic case, the 
only symmetries of that theory are those generated by the $N=4$ SCA.  
So we content ourselves with simple embedding $N=2$ SCA in the $N=4$
SCA as defined in \eqr{SimpleEmbeddingNTwoInNFourCompact} and directly 
use the result thereof, i.e. the space of gauging parameters is 
\beq    \label {eq:ParameterSpaceKThree}
    (RP^2 \times RP^2) 
\eeq
with no further reduction.  Here we are taking account of both the
left and right movers .  This space represents two distinct points on
$RP^2$ and is applicable for a generic value of the B field.  For
special values of the $B$ field the worldsheet theory also preserve
worldsheet parity and we should quotient \eqr{ParameterSpaceKThree} by
the $\ZZ_2$ that exchanges the two $RP^2$.  It represents two identical
points of $RP^2$. This
result applies not just to $\sigma$-model on generic $K3$, but also to
any $N=4$ SCFT background for $N=2$ $\beta$ string that has no additional symmetries.

\paragraph {Isometry}
What about a target space with isometry?  The isometries lead to
additional symmetries of the theory that involve the coordinate fields
$X^I$.  While some of them may commute with the $N=4$ SCA's, if there
is any that does not, it will relate two different embeddings of $N=2$
SCA.  It cannot be an inner automorphism of the $N=4$ SCA, because the
latter only affect the fermions.  So they have to quotiented out
from  the result obtained above.  One
possible scenario would be that the target space has an $SU(2)$
isometry, which corresponds to the a diagonal action of the left and
right $N=4$ SCA's $SU(2)$ outer-automorphisms.  This would reduce the
space of choices to just one $RP^2$.  
This result holds whether or not the CFT is parity invariant 
because there are two elements of $SU(2)$ that acts the the same way as 
the $\ZZ_2$ exchange symmetry: exchanging $\cal J$ and $ \tilde\cal J$.

Examples of hyper-K\"{a}hler manifolds with continuous isometries 
can be found in
\cite{Gibbons:xm}.  For example both the ALE spaces and multi-Taub-NUT
spaces have self-dual Riemann tensor and have a large isometry group
that is $SU(2) \times U(1)$.  To carries out the above procedure one
needs to identity the algebra of the isometry induced symmetries of
the field theory and the $N=(4,4)$ supercharges.  This is currently
under study.

\paragraph {$(2,2)$ Spaces}
Curved pseudo-hyper-K\"{a}hler spaces, i.e. one with metric of $(2,2)$
signature leading to a self-dual Riemann tensor, is also worth
considering.  In this paper we content ourselves with outlining the
general procedures rather than giving specific examples. If such a space
has no isometry and the holonomy group is not further reducible, the 
corresponding SCFT is likely to have no symmetry other than $N=4$ SCA.
We can then carry over the analysis for a generic $K3$
and replacing 
$SU(2)$ by $SL(2,\RR)$ in the discussion.
Again, we will content ourselves with simple embedding compact 
$N=2$ SCA in the noncompact $N=4$ SCA.
The space of gauging parameters is thus 
\beq
        (S^2_{1+} \times S^2_{1+}) \ .
\eeq
Here we are taking account of both the
left and right movers.  When the CFT has parity symmetry, we should
also quotient by it.  The result is 
\beq
        (S^2_{1+} \times S^2_{1+}) / \ZZ_2 \ .
\eeq
If there is isometry, we have to mod out its action as well.

\section{T-duality and $\alpha \leftrightarrow \beta$ Conversion}
\label {sec:TDuality}

In this section we show as promised that 
$N=2$ $\alpha$ and $\beta$ strings are related by    
 T-duality.  

\subsection{Flat Space}

T-duality and its effect on the target space geometry 
is well-known.  
Here we are interested in its relevance to $N=2$ strings.  
First we note that as T-duality is a relabeling of fields 
that relates one CFT to two different looking $\sigma$-models, 
it obviously has to exist for toroidally compactified $N=2$ strings.
It can be thought as the following field redefinition\footnote{
        For a quick look at the facts of T-duality used here, see 
    for example lecture two of \cite{Ooguri:1996ik}.
    For a comprehensive review of of T-duality, 
        see for example  \cite{Giveon:1994fu}.}
\beqar  \label {eq:TDuality}
        \pa X'^I = \pa X^I, &\csp& 
                \bar \pa X'^I = {\cal R^I}_J \bar \pa X^J \ ; \nono
        \Psi'^I = \Psi^I, &\csp& 
                \bar \Psi'^I = {\cal R^I}_J \bar \Psi^J \ .
\eeqar
For consistency with the Virasoro algebra, the OPE of $\bar\pa X'
\bar\pa X'$ and $\bar\Psi' \bar\Psi'$ must not change.  Hence $\cal R$
must be an element of the orthogonal group $O(4)$ or $O(2,2)$ 
\cite{Ooguri:1996ck}.  If we
want to keep the same $N=2$ string theory, the (super)currents we gauge
must not change.  However, $\tilde J$ now has a different expression in
terms $\bar\pa X'$ and $\bar\Psi'$ while $J$ retains the same form:
\beq
        \tilde J 
        = \frac \imath 2 {\tilde \cal J^I}_J \nod {\Psi_I \Psi^J}
        = \frac \imath 2 {\tilde \cal J'^I}_J \nod {\Psi'_I \Psi'^J}
\eeq
where $\tilde \cal J'$ is related to $\tilde \cal J$ by the rotation $\cal R$:
\beq
       \tilde \cal J' = \cal R^\top \tilde\cal J \cal R \ .
\eeq
$\cal R$ is further constrained by modular invariance that is
determined by the charge lattice of momenta and winding numbers, but
for any toroidal compactification it is nontrivial and in particular
contain elements with determinant $-1$.

Such elements are of special interest to $N=2$ strings, because as we 
have discussed in \secr{HermitianStructureEuclidean} and 
\secr{HermitianStructureTwoTwo}, they map complex structure from one
$S^2$ or $S^2_1$ to the other.  Therefore they exchange the $\alpha$
and $\beta$ types of $N=2$ strings.  The simplest example is compactification
of $X^4$ on $S^1$.  With 
\beq
        \cal R = \left( 
        \bary {cccc} 1&0&0&0\\0&1&0&0\\0&0&1&0\\0&0&0&-1 \eary\right)\ .
\eeq
representing the T-duality on the $S^1$,  $\tilde \cal J^{[3]}$ is mapped to
$tilde \cal J'^{[3]}$.  With more work, one can work out how the parameters
for $\alpha$ and $\beta$ strings are mapped into each other for all
toroidal compactification and all T-duality operations.

\subsection {Curved Space}

Flat space is of course very special.
We would like to know if the T-duality relation between $\alpha$ and 
$\beta$ strings also holds for curved spaces.  
What is needed is a version of T-duality that also works for curved 
spaces.
Fortunately, this had already been studied 
in the literature and it is worthwhile to invoke and 
reexamine these known results 
in the context of the spacetime understanding we have developed 
in the this work.  
They are couched in the language of $N=2$ superspace.  
This formalism provides a compact and efficient way to write down an $N=2$ 
$\sigma$-model, although it is not yet known that it is versatile
enough  to cover all possible cases.

We shall not repeat here the details of $N=(2,2)$ superspace in 2d.  
See for example \cite{Gates:1984nk} for a summary.
It has two copies of the same superalgebras, one  for the left and one
for the right movers.  Denote the holomorphic super-derivatives for the
left and right movers by  $D$ and $\tilde D$ respectively. 
The antii-homolomprphic ones are $\bar D$ and $\tilde {\bar D}$.
Since $\{D,\tilde D\} = \{D, \tilde {\bar{D}}\} = 0$,
in addition to the (anti-)chiral superfield $\Phi$($\bar \Phi$)\footnote{
        We shall not need their component forms, 
        which are of course expansions 
        in powers of two real Grassmann coordinates.}
\beq
\bar{D} \Phi = 0 \ , \hspace{2cm} \tilde {\bar{D}} \Phi = 0 \ , 
\eeq
we have the twisted (anti-)chiral superfield $X$($\bar X$)
 that satisfies
\beq
\bar {D} X = 0 \ ,\hspace{2cm} \tilde D X= 0 \ . 
\eeq

It is easy to see that the T-duality along, say, $\Phi + \bar\Phi$,
turns a chiral superfield into a twisted chiral superfields and 
an anti-chiral superfield into a twisted anti-chiral superfields \cite{ Buscher:sk}.  
What this means in terms of target space geometry is that, relative to
the left mover, the complex structure of the right mover changes 
sign, so the notion of (anti-)chirality is inverted for the superalgebra 
of the right mover.  Now put it in the context of a $N=2$ string theory.  
The critical dimension is $4$.  Let us start with $\beta$ theory, with 
both complex structures being, with $O(4)$ vector indices, 
\beq
     \cal J = \left(\bary {cc} 
       \imath \sigma^2 &0\\0&\imath \sigma^2 
       \eary\right) \ .
\eeq
So we have two pairs of chiral and anti-chiral multiplets.
After a T-duality transformation along, say, the 4th coordinate, 
we would instead find one chiral and anti-chiral pair 
plus one twisted and anti-twisted pairs.  
Therefore the T-duality affects the complex structure.  
The left mover should have the same complex structure for as before, 
$\cal J$, since T-duality does not affect the left movers.
Therefore the complex structure for the right mover is now 
\beq
     \tilde \cal J = \left(\bary {cc} 
       \imath \sigma^2 &0\\0&-\imath \sigma^2
       \eary\right) \ .
\eeq
It is easily verified that $\cal J$ and $\tilde \cal J$, which commute, 
resides on the two different sphere of complex structures respectively.
In the above we have used complex structure for Euclidean space, but it 
is easy to verify that the same applies to $(2,2)$ spaces.

It is a general fact in 4d that if two different complex structures commute 
with each other, they belong to two different spheres or hyperboloids 
of complex structures.  This following from 
(eq.\ref{eq:HomogeneousRelationOfComplexStructureJSUTwo}, 
\ref{eq:AlgebraicRelationOfComplexStructureJSUTwo}
\ref{eq:HomogeneousRelationOfComplexStructureJSLTwo},
\ref{eq:HomogeneousRelationOfComplexStructureJSLTwo}).
In the above example, $\beta$ string with identical complex structures 
for the left and right movers is turned into $\alpha$ strings.  
However, $\beta$ string can be more general in that the two complex structures 
may be different while lying on the same sphere or hyperboloid.  
It follows from the same set of equations in \secr{GeometryOfRFour} that
they do not commute.  Indeed it is clear that, if $\cal J$ and $\tilde \cal J$
are on the same sphere/hyperboloid but $\cal J \neq \tilde \cal J$, 
$[\cal J, \tilde \cal J]$ is non-degenerate.  Such cases, to the extent 
an $N=2$ superspace representation is known, use exactly one of 
what has been called a semi-chiral multiplet \cite{Buscher:uw}.
It is also known that a semi-chiral multiplet is related to 
one (anti-)chiral and one (anti-)twisted chiral multiplet 
by a T-duality \cite{Ivanov:ec}.  
This completes the picture of the T-duality relation between 
$\alpha$ and 
$\beta$ strings whenever a $N=2$ superspace representation is known.

The above discussion may seem somewhat abstract.  
Now we draw from the literature \cite{Buscher:sk} 
a concrete example of how 
this T-duality would affect the $N=2$ $\sigma$-model formulate 
in $N=2$ superspace.
Consider the $\beta$ theory and the action for an $N=2$ $\sigma$-model
\be   \label{eq:Bs}
        S_B = \int {d^2 x d^2} \theta d^2 \bar\theta K(\phi^1, \phi^2, 
        \bar\phi^1, \bar\phi^2)\ ,
\ee
where $\phi^i, i=1,2$ are two chiral superfields.  The real function
$K$ is called the K\"{a}hler potential and the K\"{a}hler metric is given by
\be    \label{eq:G}
G_{i\bj} = \frac{\pa^2 K}{\pa\phi^i\pa\bar{\phi}^{\bj}} \ .
\ee
If we gauge the $U(1)$ R symmetry chosen by this representation 
we would have a $\beta$ string with $\cal J = \tilde \cal J$.

Suppose now that the metric \eqr{G} admits an action of a holomorphic 
isometry.  Then there exists a local holomorphic coordinate system in which
\beq            \label{eq:KI}
 K(\phi^1, \phi^2, \bar{\phi}^1, \bar{\phi}^2) =
 K(\phi^1 + \bar{\phi}^1, \phi^2, \bar{\phi}^2) \ ,
\eeq
with the isometry generated locally by $\pa/\pa(\phi^1 - \bar{\phi}^1)$.
We can write the action \eqr{Bs} as
\beq            \label{Bss}
S = \myI {d^2 \sigma d^2 \theta d^2 \bar{\theta}}\left( K(V, \phi^2, \bar{\phi}^2) - (\chi+
\bar{\chi})V \right) \ ,
\eeq
where $V$ is an unconstrained real superfield and $\chi$ is a twisted chiral superfield.
Solving the field equations for $\chi$ and $\bar{\chi}$ yields $V=\phi^1 +\bar{\phi}^1$,
which brings us back to the action (\eqr{Bs}) with the K\"{a}hler potential \eqr{KI}.
On the other hand,   solving the field equation for $V$ yields a dual action
\beq            \label{Bss1}
\hat{S} = \myI {d^2 \sigma d^2 \theta d^2 \bar{\theta}} \left(
\hat{K}(\chi+ \bar{\chi}, \phi^2, \bar{\phi}^2)\right) \ ,
\eeq
where $\hat{K}$ is the Legendre transform of $K$ with respect to
$\phi^1 + \bar{\phi}^1$.
The function $\hat{K}$ encodes the geometry of the new target space.
The new metric $\hat{G}$ is obtained from $\hat{K}$ via
\beq            \label{gh}
\hat{G}_{1\bar{1}} = -\frac{\pa^2 \hat{K}}{\pa\chi \pa\bar{\chi}} \ ,~~~~~~~
\hat{G}_{2\bar{2}} = \frac{\pa^2 \hat{K}}{\pa\phi^2 \pa\bar{\phi}^2} \ ,
\eeq
and the other components are zero.
The new background is conformally invariant at one-loop. It has a $B$ field
\beq            \label{B}
B_{1\bar{2}} = \frac{\pa^2 \hat{K}}{\pa\chi \pa\bar{\phi}^2} \ ,~~~~~~~
B_{2\bar{1}} = \frac{\pa^2 \hat{K}}{\pa\phi^2 \pa\bar{\chi}} \ ,
\eeq
and the other components are zero.
In addition there is a dilaton $\varphi$ given by
\beq
\vp = \frac{1}{2} \log \hat G_{1\bar{1}} \ .
\eeq

The geometry of this space has been studied in  \cite{Gates:1984nk}.
It is a hermitian space which is locally a product.
It has two commuting complex structures $J^{\pm i}_{j}$.
They are covariantly constant with respect to to a connection with torsion $H=dB$,
$\Gamma^{\pm a}_{bc} = \Gamma^{a}_{bc} \mp H^{a}_{bc}$
\beq
\nabla^+J^+ = \nabla^-J^- = 0 \ .
\eeq

This establishes the T-duality at the level of the $\sigma$-models.  
Conformal invariance introduces further constraints.  The relevant 
fact is that for both type of actions it is possible.  For models 
written in twisted chiral superfields, there must be dilaton background 
if the target space is to be non-flat, while for models written in terms 
of chiral superfields, the target space can be any hyper-K\"{a}hler manifold 
without the need of a nontrivial dilaton background.  
They characterize the known examples of curved spaces for $N=2$ $\alpha$ and
$\beta$ strings  to propagate in.  It would be interesting to investigate 
these geometrical approaches further.

\section{Boundary Conditions and D-branes}    

In the previous sections we studied in detail the two families 
of $N=2$ strings $\alpha$ and $\beta$.
A natural question to ask is for 
the classification of boundary conditions
for the $N=2$ string and their geometrical interpretation.
These boundary conditions define D-branes of the $N=2$ string theories.
In this section we will discuss the subject briefly deferring the detailed
analysis to \cite{Gluck:2003pa,Yin:2002wz}\footnote{$N = 2$
 boundary conditions for non-linear $\sigma$-models have been discussed recently in 
\cite{Lindstrom:2002jb}.}.

A boundary breaks half of the superconformal symmetry.
We require the gauge
symmetry to be preserved by
the boundary condition.  The preserved half is a linear
combination of the left and right chiral currents $L^m$:
\beq
       \left. L^m ~=~\, {U^m}_n ~\tilde L^n ~~~\right|_{\pa \Sigma} \ .
\eeq
Here $L^m$ correspond to the different generators of the
chiral algebra.
The allowed gluing coefficients
${U^m}_n$ are constrained by the algebra.  Since the left and right chiral
currents have the  same algebra, consistent gluing condition must be in
one-to-one correspondence with the automorphism group of the algebra.
However, inner automorphism  can be reduced to the identity by the symmetry of
theory.  They do not lead to inequivalent boundary
conditions. Therefore we only have to consider the outer automorphisms.

One nontrivial outer automorphisms of $N=2$
SCA is the $\ZZ_2$ conjugation automorphism 
\beq    \label {eq:Conjugation}
        T \to T, \csp J \to -J, \csp, G^\pm \to G^\mp \ .
\eeq

In \cite{Ooguri:1996ck}, this $\ZZ_2$ was
used to classify the boundary conditions into two types: A-type and B-type.
For the A-type boundary condition (in the closed string notation): 
\beq \label {eq:AType}
        (L_n - \tilde L_n) \BSk_A = (J_n - \tilde J_n) \BSk_A =
        (G^+_n - \imath \tilde G^-_n) \BSk_A =  (G_n^- - \imath \tilde G_n^+)
        \BSk_A
        = 0 \ ,
\eeq
where $\BSk_A$ denotes an A-type boundary state.
For the B-type boundary condition:
\beq \label {eq:BType}
        (L_n - \tilde L_n) \BSk_B =
        (J_n + \tilde J_n) \BSk_B =
        (G^+_n - \imath \tilde G^+_n)\BSk_B=(G^-_n - \imath \tilde G^-_n) \BSk_B =
        0 \ .
\eeq
Recall that in \cite{Ooguri:1996ck} one 
arrives at these boundary conditions by first requiring that a linear
combination of $T$ and $\tilde T$ be conserved \emph {separately}, as
well as linear combination of $G$ and $\tilde G$, where $G = G^+ +
G^-$.  However, in $N=2$ string theory
there is no significance for this particular linear
combination of supercharges, since it can be rotated to
others by $SO(2)$ rotations. Also for the $N=2$ string theory 
$T$ and $J$ are on exactly the same
footing, and arbitrary linear combinations thereof, are gauge
symmetries. Therefore one can look for more general boundary conditions
that mix the two.  
In \cite{Yin:2002wz}, we discuss this in detail and arrive arrive at other
boundary conditions.

In flat space the $\sigma$-model is free and the boundary conditions can
be solved exactly. We use the free field representation of the
$N=(2,2)$ algebra.
Consider for instance the $B$ boundary states.
Equations  \eqr{BType} can be solved by 
\beq
        (a^i_m - {R^i}_j \tilde a^j_{-m}) \BSk_{B} \csp 
        = \csp (a^{\bar i}_m 
        - {R^{\bar i}}_{\bar j} \tilde a^{\bar j}_{-m}) \BSk_{B} \csp 
= \csp 0 \ ,
\label{eq:geom}
\eeq
and 
\beq
        (\Psi^i_m - \imath {R^i}_j \tilde \Psi^j_{-m}) \BSk_{B} \csp
        = \csp (\Psi^{\bar i}_m - 
                \imath {R^{\bar i}}_{\bar j} \tilde \Psi^{\bar j}_{-m}) 
                        \BSk_{B} \csp
        = \csp 0 \ ,
\eeq
where $a^i_m$ and $\Psi^i_m$ are the worldsheet bosonic and fermionic
oscillators.
The matrix $R$ satisfies 
\beq
        ({R^i}_j)^* = {R^{\bar i}}_{\bar j} \ .
\eeq
For $(4,0)$ signature $R$ is an $U(2)$ matrix, while 
for a $(2,2)$ signature it is a $U(1,1)$ matrix.
Note that equations \eqr{geom} 
determine the geometry and the (open string) gauge field of the D-brane, as
\beq
        (\pa X^i - {R^i}_j \bar\pa X^j) \BSk_{B} \csp 
        = \csp (\pa X^{\bar i} - 
        {R^{\bar i}}_{\bar j} \bar \pa X^{\bar j}) \BSk_{B} \csp
        = \csp 0 \ .
\eeq
In the absence of gauge fields, vectors of $R$ with eigenvalues
$(-1)$ and $(+1)$ correspond to directions normal and tangential 
to the D-brane, respectively. 
Examples of boundary state that satisfies the above conditions
 can be easily constructed \cite{Yin:2002wz}.
The type $A$ works in a similar way.
One way is to perform a mirror symmetry on the right movers.
Instead of ${R^i}_j$ and ${R^{\bar i}}_{\bar j}$, one would have
$R^{i}_{\bar j}$ and ${R^{\bar i}}_{j}$, which again must be
unitary or pseudo-unitary.

In \cite{Gluck:2003pa} we discuss the geometrical meaning of the
boundary conditions when the target space is curved.
The boundary conditions correspond to even-dimensional D-branes for
the type $\beta$ family and odd-dimensional for type $\alpha$.
They are mapped to each other under T-duality.

\section*{Acknowledgments}

EC would like to thank A.~Brandhuber, A.~Kapustin, C.~Vafa, and J.~Schwarz  for discussion; and Perimeter Institute,  Institute for Advanced Studies, and CERN, 
where parts of the work were done, for hospitality.  
ZY would like to thank C.~Hull for discussion and references,
E.~Martinec for
discussion, L.~Alvarez-Gaume, J.~Barbon, W.~Lerche for comments, 
H.~Ooguri for originally getting him interested in the
subject of $N=2$ strings, and Isaac Newton Institute for Mathematical
Sciences for hospitality.

EC is supported by John A. McCone Foundation and by Department of Energy grant
DE-FG03-92-ER40701.
The research of YO is supported by the US-Israel Binational Science Foundation.

\newpage
\appendix

\section* {APPENDIX}     

\section {Summaries of Notations}

\subsection {Symbols and Indices}

\indent 

$\RR^4$ vector: $I,J \ldots \in \{0,1,2,3\}$.

$C^2$ complex coordinates: $i,j\ldots; \bar i \bar j \ldots \in \{1,2\}$

Affine $SU(2)$ and $SL(2,\RR)$ currents:
$i,j,\ldots \in \{1,2, 3, \}$

(Bi-)spinor, for $SU(2)$ and $SL(2,\RR)$: $\alpha,\beta,\ldots \in \{1,2\}$.

\subsection {Worldsheet Geometry}

\indent

$\Sigma$: Worldsheet manifold.

$\pa$, $\tilde \pa$:  Spacetime derivatives for the left movers and right movers respectively.

$\pa$, $\bar\pa$:       Worldsheet derivatives for the left and right movers respectively.

$D$($\tilde D$), $\bar D$($\bar {\tilde D}$):  Left (right) holomorphic/anti-holomorphic superderivatives in $N=2$ superspace.

\subsection {Worldsheet Matter Fields}

\indent

$X^I$: bosonic fields mapping worldsheet into target space.

$\Psi^I$: fermionic fields mapping worldsheet into target space tangent bundle.

$\bX^I$: worldsheet superfields.  $\bX^I = X^I + \theta \Psi^I$.

\subsubsection {Affine Currents}

\indent

$T$: stress tensor (as a local field); 
$\csp L_m$: Virasoro operator (as mode operator)

$J$($\tilde J$): Left (right) affine $U(1)$ current of $N=2$ SCA.

$G^\pm$($\tilde G^\pm$): Left (right) affine supercurrents of $N=2$ SCA

$J^i$($\tilde J^i$):   Left (right) affine $SU(2)$ or $SL(2,\RR)$ currents

$G^{\alpha,\beta}$ ($\tilde G^{\alpha,\beta}$):  Left (right) affine supercurrents of $N=4$ SCA

\subsubsection {Target Space Geometric Data}

\indent

$\fX$: target space manifold

$\cal G$: metric

$\cal J$: complex structure. $\cal J \cal J = -1$.

$\cal K$:  K\"{a}hler form. $\cal K = \cal G \cal J$.

$\cal B$: Rank 2 Anti-symmetric tensor

$\cal H$: The field strength for $\cal B$: $\cal H = d \cal B$.

$D$: covariant derivative

$\Gamma$: affine connection

$\Gamma_0$: Levi-Cevita connection based on $\cal G$.

\subsubsection {Algebraic Invariants}

\indent

$\epsilon^{ijk}$: $su(2)=so(3)$ structure constants.

$\delta^{ij}$:  $su(2)=so(3)$ invariant Kronecker $\delta$ tensor.

${(\sigma^i)^\alpha}_\beta$: $su(2)$ invariant Pauli matrices.

$\varepsilon^{ij}_k$: $sl(2,\RR)=so(2,1)$ structure constants.

$\eta^{ij}$:  $sl(2,\RR)=so(2,1)$ Killing metric.

\subsubsection {Geometric Objects}
\label  {sec:GeometricObjects}

\indent

$S^2$: 2-sphere, $\{(a_1, a_2, a_3) | a_1^2 + a_2^2 + a_3^2 = 1 \}$.

$RP^2$: real projective sphere, $S^2 / (\vec a \sim - \vec a)$.

$S^2_1$: hyperboloid of two sheets, 
      $\{(a_1, a_2, a_3) |a_1^2 - a_2^2 + a_3^2 = -1 \}$.

$S^2_{1+}$: Upper sheet of hyperboloid of two sheets, 
      $s^2_1 \cap \{(a_1, a_2, a_3) | a_2 > 0\}$.

\subsection {Conventions}

\indent

\beq
       \epsilon^{123} = 1
\eeq

\beq
        \epsilon_{\alpha\beta} 
        = \left(\bary {cc} 0&-1\\1&0 \eary\right), \csp
        \epsilon^{\alpha\beta} = 
        \left( \bary {cc} 0&1\\-1&0 \eary\right) \ .
\eeq

\beqar
{(\sigma^1)^\alpha}_\beta &=& \left(\bary {cc} 0&1\\1&0 \eary\right),\nono
{(\sigma^2)^\alpha}_\beta &=& \left(\bary {cc} 0&-\imath\\\imath&0 \eary \right),\nono
{(\sigma^3)^\alpha}_\beta &=& \left(\bary {cc} 1&0\\0&-1 \eary\right) \ .
\eeqar
\beq
        \sigma^{\pm} = \sigma^1 \pm i \sigma^2 \ .
\eeq

\beq
        (\sigma^i)_{\alpha\beta} 
        = \epsilon_{\alpha\gamma} {(\sigma^i)^\gamma}_\beta, \csp 
        (\sigma^i)^{\alpha\beta} 
        = {(\sigma^i)^\alpha}_\gamma \epsilon^{\gamma\beta}\ .
\eeq

\beq    \label {eq:etaConvention}
        \eta^{ij} = \left(\bary {ccc} 
        -1&0&0\\0&-1&0\\0&0&1 
        \eary \right) \ .
\eeq

\beq    \label {eq:epsilonConventionTwoOne}
        \varepsilon^{12}_3 = \varepsilon^{23}_1 = \varepsilon^{13}_2 = 1 \ .
\eeq

\section {Hyperboloid}
\label {sec:Hyperboloid}

Hyperboloid of two sheets is the described 
by the solution in $\RR^3$ of the equation 
\beq
    x^2 + y^2 - z^2 = -1
\eeq
It is so named because, obviously, it has two components (sheets), 
corresponding respectively to positive and negative values of $z$.  
It is known as the pseudo-2-sphere of index 1 with imaginary radius: 
$S^2_1$ \cite{soviet}.  
The upper sheet, or equivalently the $\ZZ_2$ quotient by $z \to -z$, 
of $S^2_1$ is denoted by $S^2_{1+}$ in this paper.

One effective way to parameterize $S^2_{1+}$ is to use $x$ and $y$.  
Hence there is a one-to-one map between $R^2$ and $S^2_{1+}$.
Thus the polar coordinates $\theta$ and $r$ of $\RR^2$ 
is equally good for $S^2_{1+}$.

$O(2,1)$ acts on $S^2_1$ just naturally as $O(3)$ acts on $S^2$.  
The connected subgroup of $O(2,1)$ is $SO_0(2,1)$.  
It can be generated by a rotation of of the xy-plane with angle $\phi$
around the $z$ axis, which acts by
\beq
    r \to r , \csp \theta \to \theta + \phi,
\eeq
and a Lorentz boost along the $x$-direction with $z$ as the ``time.'' 
It acts transitively on plane sections of $S^2_1$ with fixed value of $y$.

\newpage


\end{document}